\documentclass[a4paper,11pt]{article}

\usepackage[T1]{fontenc}

\usepackage{mathtools} 
\usepackage{amssymb}
\usepackage{braket}

\usepackage{graphicx}
\usepackage{xcolor}
\usepackage{microtype}  
\usepackage{booktabs}   
\usepackage{subcaption} 
\usepackage[percent]{overpic} 
\usepackage{aas_macros} 
\usepackage{comment}
\usepackage{tikz}
\usetikzlibrary{
  matrix,
  fit,
  positioning,
  arrows.meta,
  decorations.pathreplacing,
  shapes.geometric,
  backgrounds
}

\usepackage{jcappub}

\newcommand{\vk}{\mathbf{k}}
\newcommand{\vq}{\mathbf{q}}
\newcommand{\vx}{\mathbf{x}}
\newcommand{\vPsi}{\mathbf{\Psi}}
\newcommand{\vphi}{\phi}
\newcommand{\vPhi}{\Phi}

\title{\boldmath Mapping the Evolving Cosmic Web on the Lightcone with
{WebON}: {HOLI} Mock Strategy}

\author[1,2]{Francisco-Shu Kitaura,}
\author[3,4,5,6,1,2]{Francesco Sinigaglia,}
\author[1,2]{Kilian Peraza Vera,}
\author[1,2]{Pere Rosselló,}
\author[7]{Daniel Forero-Sánchez,}
\author[8]{Natalia Villa-Nova Rodrigues,}
\author[9]{Jorge E. Garc\'ia-Farieta,}
\author[10]{Yunyi Tang,}
\author[11]{Jose María Coloma-Nadal,}
\author[1,2]{Ginevra Favole,}
\author[1,2]{Aurelio Carnero-Rosell,}
\author[10]{Cheng Zhao,}
\author[12,13,14,15,16]{Kentaro Nagamine}

\emailAdd{fkitaura@iac.es}
\emailAdd{fsinigag@sissa.it}

\abstract{
Wide-area spectroscopic surveys such as DESI require large ensembles
of mock catalogues that follow gravitational evolution and tracer
populations across the observer's past lightcone. We present
\texttt{WebON} (Web ON the lightcone), a fast method for mapping the
evolving dark-matter cosmic web onto the lightcone without periodic-box
replication or the stitching of separately evolved snapshot shells.
Displacements computed at a limited set of redshifts are combined with
a particle-wise Zel'dovich estimate of the emission redshift and a
continuous-redshift correction, allowing lightcone positions to be
generated in a single forward pass. The method supports regularised
ALPT and eALPT gravity solvers, with an optional short-range ridging
correction. We illustrate full-sky matter and cosmic-web evolution and
find percent-level agreement between lightcone and fixed-redshift angular
power spectra over the tested redshift and multipole ranges, providing
an internal consistency check of the lightcone construction. Optional
modules address approximate relativistic position corrections and
local primordial non-Gaussianity. Finally, we show how the spectral cosmic-web hierarchy connects
\texttt{WebON} to the \texttt{HOLI} (Higher Order LIghtcone) mock
strategy, in which post-processing ridging and calibrated
\texttt{HICOBIAN} bias and subgrid prescriptions can recover tracer
clustering beyond the nominal resolution of the ALPT matter field.
Independent comparisons of ridged LPT matter fields with Abacus
$N$-body simulations and published percent-level tests of the
hierarchical tracer-bias model complement the lightcone consistency
measurements presented here.
\vspace{1cm}\\\noindent{\small\itshape
Author affiliations are listed at the end of the manuscript.}}

\begin{document}

\maketitle

\flushbottom

\section{Introduction}
\label{sec:intro}

The Dark Energy Spectroscopic Instrument (DESI) is mapping large-scale
structure over an unprecedented combination of comoving volume and
redshift range, providing precision baryon acoustic oscillation (BAO)
and redshift-space distortion (RSD) measurements from multiple tracer
populations. Robust inference requires mock catalogues that capture
gravitational evolution, tracer clustering, and survey selection, in
numbers sufficient for covariance estimation and systematic tests.

High-resolution $N$-body simulations accurately model nonlinear
structure formation but face a demanding volume--resolution--cost
trade-off when thousands of DESI-scale realisations are required.
Fast approximate gravity solvers coupled to calibrated field-level
tracer models offer a complementary approach: large-scale matter
evolution is computed explicitly, while unresolved tracer physics is
represented statistically.

The cosmic web---the network of knots, filaments, sheets, and voids---
provides a natural connection between gravity and tracer formation
\citep[e.g.,][]{2009LNP...665..291V}. It is relevant to baryons in
filaments \citep[][]{2015ApJ...806..113G,2015Natur.528..105E,2019A&A...624A..48D},
BAO analyses using voids
\citep[][]{2016PhRvL.116q1301K,2016MNRAS.459.4020L,2020MNRAS.491.4554Z},
RSD constraints \citep[][]{2017PhRvD..95f3528C,nadathur18},
tests of dark energy and modified gravity
\citep[][]{2007PhRvL..98h1301P,2010MNRAS.403.1392L,2012MNRAS.426..440B,2015PhRvD..92h3531P,2019MNRAS.484.5267K},
and galaxy formation as a function of environment
\citep[e.g.,][]{2001ApJ...557..117B,2004ApJ...615L.101B,2004MNRAS.353..713K,2005ApJ...629..143B,2006MNRAS.373..469B,2007ApJ...655L...5A,2008ApJ...688...78L,2008MNRAS.383..907B,2012MNRAS.421L.137L,2013AJ....145..120B,2014MNRAS.445..988N,2015ApJ...802...82F,2017ApJ...848...60Y,2018MNRAS.473.2486S,2018MNRAS.473.3941F,2019MNRAS.485.2367C,2019MNRAS.483.4501A,2019OJAp....2E...7A,2019MNRAS.482.1900H,2019MNRAS.487.1607G}.
Different web classifiers emphasise different aspects of the matter
and velocity fields \citep[][]{2018MNRAS.473.1195L}; for tracer
modelling, this is useful rather than merely ambiguous because
classification at different scales captures complementary nonlocal
information.

The underlying dynamics can be formulated through the
Lagrangian--Eulerian map
\begin{equation}
\vx(\vq,z)=\vq+\vPsi(\vq,z),
\label{eq:lag_eul_map}
\end{equation}
where $\vPsi$ is the displacement field
\citep[see][for a review]{2002PhR...367....1B}. In potential-flow
approximations, $\vPsi=-\nabla\vPhi$, and potential Hessians
describe anisotropic gravitational collapse. The tidal tensor
$\mathcal{T}_{ij}\equiv\partial_i\partial_j\vphi$,
with $\nabla^2\vphi=\delta$, provides a familiar web descriptor
\citep[e.g.,][]{1970A&A.....5...84Z,Bond_1996nat,2007MNRAS.375..489H}.
 However, tracer bias involves nonlinear and nonlocal contributions
(beyond local-in-matter-density bias) that cannot be fully captured
by any single tidal classification
\citep[see][and references therein]{2018PhR...733....1D,McDonald09,2012PhRvD..85h3509C}.
A spectral hierarchy of web operators can encode tidal and
higher-derivative information in a compact environmental
representation \citep[][]{Coloma_2024,2026arXiv260315834K}.
Indeed, nonlinear and nonlocal bias mappings incorporating cosmic-web
dependencies have been shown to emulate the effects of modified gravity
on tracer clustering, even when applied to standard $\Lambda$CDM
density fields \citep[][]{García-Farieta2024}.

The main contribution of this work is \texttt{WebON}, which assigns
particle emission redshifts and evaluates approximate gravitational
displacements directly on a continuously evolving dark-matter
lightcone. We present its numerical construction and examine the internal
consistency of the resulting matter and cosmic-web fields as they
evolve along the lightcone. We then describe
how those fields enter \texttt{HOLI} (Higher Order LIghtcone mocks),
through spectral web classification and the environment-conditioned
\texttt{HICOBIAN} bias model \citep[][]{Coloma_2024}. The present figures focus on the internal consistency of the
\texttt{WebON} lightcone construction. Independently, comparisons of
ridged LPT matter fields against Abacus $N$-body simulations
\citep{2026arXiv260313106K} and the demonstrated accuracy of
\texttt{HICOBIAN} halo clustering on coarse ALPT fields
\citep{Coloma_2024} establish the performance of complementary
components. Their combination and extension to tracer-specific
lightcone catalogues are the subject of companion studies
(Forero-S\'anchez et al. and Tang et al., in preparation).

Section~\ref{sec:webon} introduces \texttt{WebON} and its optional
extensions. Section~\ref{sec:bias} connects its cosmic-web fields
to the \texttt{HOLI} mock strategy. Section~\ref{sec:validation}
presents lightcone, angular-clustering, and large-scale consistency
tests. We conclude in Sec.~\ref{sec:conclusions}.

\section{WebON: Dark-Matter Lightcone Production}
\label{sec:webon}

\texttt{WebON} constructs observer-centred dark-matter lightcones
from periodic initial conditions without replicating the simulation
volume or joining independently evolved matter shells. Its inputs
are approximate displacement fields computed at a selected set of
redshifts. The lightcone is assembled by assigning each particle an
emission redshift, correcting the stored displacement to that
redshift, and selecting its observer-centred position. The same
procedure can be used with different approximate gravity solvers.

\subsection{Displacement snapshots and lightcone crossing}
\label{sec:webon:streaming}
\label{sec:webon:redshift}

We first compute and store displacement fields
$\vPsi_{\rm app}(\vq,z_{\rm B})$ at the chosen redshift-bin
boundaries using regularised Lagrangian perturbation theory (LPT),
as well as the first-order displacement
$\vPsi^{(1)}(\vq)\equiv\vPsi_{\rm LPT}(\vq,z=0)$.
The subsequent particle loop may be parallelised in redshift-bin
chunks. At each Lagrangian coordinate $\vq$, the initial estimate
$z_q$ is defined by the observer distance,
\begin{equation}
\chi_q=|\vq-\vx_{\rm obs}|,\qquad \chi(z_q)=\chi_q,
\end{equation}
where $\chi(z)$ denotes the comoving distance--redshift relation
and $\vx_{\rm obs}$ the observer position. With the linear growth
factor normalised to $D(0)=1$, a Zel'dovich proxy gives
\begin{equation}
\vx_{\rm Zel}=\vq+D(z_q)\vPsi^{(1)}(\vq),
\end{equation}
and the approximate emission redshift follows from
\begin{equation}
\chi(z_r)=|\vx_{\rm Zel}-\vx_{\rm obs}|.
\end{equation}
This estimate incorporates coherent bulk flows without repeatedly
evaluating the full gravity solver. It does not, by itself, solve
the final approximate trajectory's lightcone-crossing condition.

For $z_{\rm B}\le z_r$ at the lower boundary of the relevant
redshift interval, we evaluate the displacement using
\begin{equation}
\vPsi(\vq,z_r)\simeq \vPsi_{\rm app}(\vq,z_{\rm B})
+\left[D(z_r)-D(z_{\rm B})\right]\vPsi_{\rm LPT}(\vq,z=0),
\label{eq:webon_continuous}
\end{equation}
with sufficiently close displacement snapshots for the linear
increment to remain subdominant. We then assign
$\vx(\vq,z_r)=\vq+\vPsi(\vq,z_r)$ and apply the lightcone
selection at $\chi(z_r)$. An optional relativistic coordinate
correction can be added as described in Sec.~\ref{sec:webon:gr}.
The residual
\begin{equation}
\epsilon_\chi(\vq)=
|\vx(\vq,z_r)-\vx_{\rm obs}|-\chi(z_r)
\label{eq:lightcone_residual}
\end{equation}
quantifies the difference between the Zel'dovich emission-redshift
proxy and the final approximate trajectory; it should be assessed
against the radial resolution required for the analysis.

\subsection{Gravity solvers, short-range completion, and velocities}
\label{sec:webon:short-range}
\label{sec:webon:velocity}

We use regularised LPT solvers at particle/mesh scales of
approximately $1$--$6\,h^{-1}\mathrm{Mpc}$, rather than attempting
to resolve individual galaxy-host haloes. Augmented LPT (ALPT)
\citep{Kitaura_2013} combines large-scale 2LPT displacements with
a spherical-collapse-inspired ultraviolet regularisation and is
used at coarser resolution ($\gtrsim 2.5\,h^{-1}\mathrm{Mpc}$).
Eulerian-ALPT (eALPT) \citep{Kitaura_2024} incorporates iterative
Eulerian corrections, including effective stress-tensor
contributions and controlled vorticity, and is used at finer
resolution ($\lesssim 2.5\,h^{-1}\mathrm{Mpc}$).

An optional short-range Eulerian ridging correction can be applied
to precomputed ALPT or 2LPT displacements
\citep[][]{2026arXiv260313106K}. This post-processing step sharpens
dense ridges without repeating the long-range gravity calculation;
its clustering performance depends on the adopted calibration.
Tetrahedral phase-space mapping can additionally reduce discreteness
artefacts in mesh-based density reconstruction \citep{Abel_2012}.

Peculiar velocities are computed in post-processing based on the (e)ALPT and LPT displacement fields, using
\begin{equation}
\begin{aligned}
\vec{v}(\vq,z)
&=
\frac{f(z)H(z)}{1+z}\left[
D(z)\vPsi_{\rm LPT}(\vq,z=0)
+\frac{f_2(z)}{f(z)}\left(\vPsi(\vq,z)-D(z)\vPsi_{\rm LPT}(\vq,z=0)\right)\right] \\
&\simeq
\frac{f(z)H(z)}{1+z}\left[
D(z)\vPsi_{\rm LPT}(\vq,z=0)
+\frac{f_2(z)}{f(z)}\left(\vPsi(\vq,z_{\rm B})-D(z_{\rm B})\vPsi_{\rm LPT}(\vq,z=0)\right)\right],
\end{aligned}
\label{eq:webon_velocity}
\end{equation}
where $f(z)\equiv d\ln D/d\ln a$ is the linear growth rate and $f_2(z)$ denotes the corresponding second-order growth-rate factor in the adopted perturbative scheme.
This construction recovers the linear large-scale velocity relation while retaining a solver-dependent approximation to nonlinear corrections. The residual displacement is assigned a second-order growth rate; for regularised or iterated solvers, this is an effective velocity prescription rather than an exact time derivative of the full trajectory.
When using eALPT, the dependence on the full displacement field permits retaining any vorticity component generated by the Eulerian extension. Here $\vPsi$ denotes the dynamical displacement, before the optional GR coordinate shift; the latter should not be interpreted as an additional physical peculiar velocity.

\subsection{Optional ultra-large-scale extensions}
\label{sec:webon:extensions}

The following modules extend the baseline lightcone calculation but
do not alter its particle-wise emission-redshift construction.

\subsubsection{Approximate relativistic position correction}
\label{sec:webon:gr}

The displacement fields used in \texttt{WebON}  are computed in Newtonian perturbation theory.
This is an excellent approximation for the DESI clustering regime, but it formally neglects relativistic constraint corrections that become relevant on ultra-large (horizon-approaching) scales.
Following {\it Hahn and Paranjapee} \citep[][]{2016PhRvD..94h3511H}, we optionally include an approximate position-level correction that maps simulation-like
particle positions to their Newtonian-gauge counterparts on
ultra-large scales.

\paragraph{Helmholtz form and screened kernel.}
\label{sec:webon:gr_helmholtz}

A convenient way to encode the relativistic constraint correction is through a Helmholtz-type modification of the Poisson relation for the potential,
\begin{equation}
\left(\nabla^2-\mu^2(z)\right)\phi(\vx,z)=\delta(\vx,z),
\label{eq:helmholtz_phi}
\end{equation}
with $\mu(z)\equiv k_{\rm GR}(z)=\ell^{-1}(z)$ a time-dependent screening wavenumber. Here the positive cosmological prefactor in the potential equation has been absorbed into $\phi$ at fixed redshift; this rescaled potential should be distinguished from the primordial potential used in Sec.~\ref{sec:webon:png}. We use the Fourier convention $\nabla\leftrightarrow i\vk$ and $\nabla^2\leftrightarrow-k^2$.
The Helmholtz construction motivates the screening scale used below.
However, \texttt{WebON} does not rerun the gravitational evolution
with the screened force kernel. Instead, it applies the approximate
position-level transformation derived by Hahn and Paranjapee 2016
\citep[][]{2016PhRvD..94h3511H} to the completed simulation-like
lightcone.
The Helmholtz relation itself reads in Fourier space
\begin{equation}
\hat{\phi}(\vk,z)=-\frac{\hat{\delta}(\vk,z)}{k^2+\mu^2(z)}\,,
\label{eq:phi_screened}
\end{equation}
so that the associated gradient kernel, written in displacement notation as $\vPsi_{\rm H}=-\nabla\phi$, becomes
\begin{equation}
\hat{\vPsi}_{\rm H}(\vk,z)=i\frac{\vk}{k^2+\mu^2(z)}\,\hat{\delta}(\vk,z)\,.
\label{eq:psi_screened}
\end{equation}

The Helmholtz kernel regularises the response of the potential on
scales comparable to the screening length. In particular, the
difference between the screened and Newtonian force displacements is
\begin{equation}
\Delta\widehat{\vPsi}_{\rm H}
\equiv
\widehat{\vPsi}_{\rm H}-\widehat{\vPsi}_{\rm N}
=
i\vk
\left[
\frac{1}{k^2+\ell^{-2}(z)}-\frac{1}{k^2}
\right]
\widehat{\delta}(\vk,z).
\end{equation}
For $k\gg\ell^{-1}$, this difference is proportional to
$-\ell^{-2}\nabla\nabla^{-4}\delta$.

This screened-force displacement is introduced only to motivate the
relativistic screening scale. It is not the correction applied in
\texttt{WebON}. The implemented correction is instead the
position-level transformation from simulation-like to approximate
Newtonian-gauge particle positions described below.

\paragraph{Implementation used in \texttt{WebON}.}
\label{sec:webon:gr_impl}

In \texttt{WebON}, we implement the approximate position-level
gauge transformation proposed by Hahn and Paranjapee 2016 \citep[][]{2016PhRvD..94h3511H} as an
additional displacement field computed on the mesh and evaluated at
the particle emission redshift. Specifically, the displacement that
maps simulation-like particle positions to their approximate
Newtonian-gauge counterparts is written as
\begin{equation}
\vPsi_{\rm GR}(\vq,z)
=
\mathcal{F}_{\rm GR}(z)\,
\nabla\nabla^{-4}\delta_{\rm sim}(\vq,z),
\label{eq:webon_gr_real}
\end{equation}
where
\begin{equation}
\mathcal{F}_{\rm GR}(z)\equiv \ell^{-2}(z)
\end{equation}
and $\ell(z)$ is the relativistic screening scale. In terms of the
conformal Hubble rate,
\begin{equation}
\mathcal{H}(z)\equiv a(z)H(z)=\frac{H(z)}{1+z},
\end{equation}
the screening length is $\ell=c/(\sqrt{3f}\,\mathcal{H})$ in the growing-mode approximation. Using the approximation for $f$ below gives
\begin{equation}
\ell(z)
\simeq
\frac{c}{\sqrt{3}\,\mathcal{H}(z)}
\Omega_{\rm m}^{-2/7}(z),
\label{eq:screening_scale}
\end{equation}
where we have used
$f(z)\simeq\Omega_{\rm m}^{4/7}(z)$.
Equivalently, this expression may be written as
\begin{equation}
\ell(z)
\simeq
\frac{c}{\sqrt{3}\,H_0}
\Omega_{\rm m0}^{-2/7}
a^{2/7}(z)
\left[\frac{\mathcal{H}(z)}{H_0}\right]^{-3/7}.
\label{eq:screening_scale_equiv}
\end{equation}

The corresponding Fourier-space displacement is
\begin{equation}
\widehat{\vPsi}_{\rm GR}(\vk,z)
=
i\frac{\vk}{\ell^2(z)k^4}\,
\widehat{\delta}_{\rm sim}(\vk,z),
\qquad k\neq0,
\label{eq:webon_gr_fourier}
\end{equation}
with the zero mode set to zero. On the ultra-large scales relevant
for this correction, where the density evolution can be approximated
by linear growth,
\begin{equation}
\delta_{\rm sim}(\vq,z)
\simeq
D(z)\,\delta_{\rm sim}(\vq,z=0),
\end{equation}
and therefore
\begin{equation}
\vPsi_{\rm GR}(\vq,z)
\simeq
D(z)\,\ell^{-2}(z)\,
\nabla\nabla^{-4}\delta_{\rm sim}(\vq,z=0).
\label{eq:webon_gr}
\end{equation}
This correction is added to the simulation-like particle position,
$\vx_{\rm N}\simeq\vx_{\rm sim}+\vPsi_{\rm GR}$, and should not be
confused with replacing the full Newtonian displacement by the
Helmholtz-screened force displacement.

\paragraph{Regime of relevance and numerical aspects.}
\label{sec:webon:gr_regime}

For standard DESI BAO and RSD analyses, the dominant modes lie well inside the horizon, where the relative correction scales as $[k\ell(z)]^{-2}$. Its size depends on both redshift and wavenumber; consequently, a single redshift-independent cutoff should not be used to define where it is negligible. Figure~\ref{fig:gr_probe} illustrates this dependence explicitly, including high-redshift tests beyond the main DESI lightcone range.
Nevertheless, including $\vPsi_{\rm GR}$ is useful for applications involving horizon-scale modes (e.g. ultra-large-scale clustering, relativistic lightcone effects, or cross-correlations with CMB/ISW-like observables), and its computational overhead is small because it requires only an additional FFT-based vector operator rather than a new gravitational evolution.

Numerically, $\nabla^{-4}$ enhances the infrared, so one must treat the $k=0$ mode consistently (set to zero in a periodic box) and adopt consistent Fourier conventions.
Because the correction is evaluated at the emission redshift $z_r$ and does not introduce additional snapshot stitching, it preserves the continuous-redshift structure of the \texttt{WebON}  construction. The comparison with the power-spectrum ratio in Fig.~\ref{fig:gr_probe} is discussed in Sec.~\ref{sec:validation:gr}. This position correction alone does not constitute a complete relativistic prediction for observed number counts: photon propagation, observed-redshift perturbations, magnification, and selection effects require a consistent treatment of light propagation and survey observables.

\subsubsection{Local primordial non-Gaussianity}
\label{sec:webon:png}

\texttt{WebON} can construct lightcones from initial conditions with
local primordial non-Gaussianity (PNG), parameterised by
$f_{\rm NL}^{\rm loc}$. The primordial potential is generated as
\begin{equation}
\Phi(\vx)
=
\phi_{\rm G}(\vx)
+
f_{\rm NL}^{\rm loc}
\left[
\phi_{\rm G}^{2}(\vx)
-
\left\langle\phi_{\rm G}^{2}\right\rangle
\right],
\label{eq:local_png}
\end{equation}
where $\phi_{\rm G}$ is a Gaussian random field. The resulting
potential is converted into the initial density and displacement
fields using the same transfer-function, growth-factor, and Fourier
conventions as in the Gaussian case. The subsequent \texttt{WebON}
lightcone construction is unchanged.

For a biased tracer, local PNG generates the large-scale response
\begin{equation}
\Delta b_{\rm PNG}(k,z)
=
f_{\rm NL}^{\rm loc}
\frac{b_{\Phi}(z)}{\mathcal{M}(k,z)},
\label{eq:png_bias_general}
\end{equation}
where
\begin{equation}
\mathcal{M}(k,z)
=
\frac{2}{3}
\frac{c^2 k^2T(k)D(z)}
{\Omega_{\rm m0}H_0^2},
\label{eq:png_transfer}
\end{equation}
for a dimensionless potential extrapolated to the present-day large-scale convention and a growth factor normalised to $D(0)=1$. If $\Phi$ and $f_{\rm NL}^{\rm loc}$ are instead defined in the primordial matter-era (CMB) convention, the right-hand side of Eq.~\eqref{eq:png_transfer} must include the factor $g(0)/g(z_{\rm md})$, where $g(z)=(1+z)D(z)$ and $z_{\rm md}$ lies in matter domination. The potential, input $f_{\rm NL}$, and reference simulations must all use the same convention.
In units with $c=1$, the explicit factor of
$c^2$ is omitted. A commonly used parametrisation of the PNG response
is
\begin{equation}
b_{\Phi}(z)
\simeq
2\delta_{\rm c}\left[b_1(z)-p\right],
\end{equation}
where $p=1$ corresponds to the universal-mass-function prediction,
while more general tracer selections may require a calibrated
$b_{\Phi}$. Since $T(k)\rightarrow1$ on sufficiently large scales,
Eq.~\eqref{eq:png_bias_general} yields the characteristic
$\Delta b_{\rm PNG}\propto f_{\rm NL}^{\rm loc}k^{-2}$ response
\citep{Dalal_2008}.

A companion study, \emph{Primordial non-Gaussianity as a large-scale
cosmic-web bias} \citep[][]{Peraza_2026}, shows that this
scale-dependent response can be reformulated as a nonlocal,
anisotropic environmental-bias effect. The construction introduces
the large-scale Hessian field
\begin{equation}
\mathcal{H}^{\rm PNG}_{ij}(\vx;R)
\equiv
\partial_i\partial_j\nabla^{-4}\delta_R(\vx),
\label{eq:png_hessian}
\end{equation}
whose trace satisfies
\begin{equation}
{\rm Tr}\,\mathcal{H}^{\rm PNG}
=
\nabla^{-2}\delta_R.
\label{eq:png_hessian_trace}
\end{equation}
Its trace therefore carries the same inverse-Laplacian, or $k^{-2}$,
enhancement that appears in the local-PNG scale-dependent bias.
Diagonalising $\mathcal{H}^{\rm PNG}_{ij}$ defines four large-scale
knot-, filament-, sheet-, and void-like environments. These should not
be interpreted as the conventional nonlinear cosmic web: they are
large-scale, nonlocal environments designed to isolate the PNG
response.

The companion analysis assigns one weight to each of these four
environments and tests the construction against the
\textsc{AbacusPNG} simulations with
$f_{\rm NL}^{\rm loc}=\{-100,-30,30,100\}$. The four environmental
weights reproduce the large-scale $k^{-2}$ rise or suppression of the
tracer bias found in the corresponding non-Gaussian simulations. They
also exhibit a robust sign-dependent hierarchy,
\begin{align}
f_{\rm NL}^{\rm loc}>0:\quad&
w_{\rm K}>w_{\rm F}>w_{\rm S}>w_{\rm V},
\\
f_{\rm NL}^{\rm loc}<0:\quad&
w_{\rm K}<w_{\rm F}<w_{\rm S}<w_{\rm V},
\end{align}
so that positive PNG preferentially enhances knot-like environments,
whereas negative PNG reverses the ordering and gives the largest
relative weight to void-like environments.

The reconstruction can be extended towards higher wavenumbers by
including the standard tidal tensor
\begin{equation}
\mathcal{T}_{ij}(\vx;R)
\equiv
\partial_i\partial_j\nabla^{-2}\delta_R(\vx).
\label{eq:png_tidal_tensor}
\end{equation}
Four additional, factorised tidal-environment weights capture the
residual anisotropic response not described by the
$\nabla^{-4}$ Hessian alone. The resulting eight-weight model
reproduces the measured PNG power-spectrum response within the
cosmic-variance uncertainty of the simulation ensemble over the
tested range.

This construction provides a complementary route to PNG lightcone
mocks. Existing Gaussian halo or galaxy catalogues can be reweighted
or downsampled according to their large-scale web environment,
imprinting an effective local-PNG scale-dependent bias while
retaining the positions and velocities of the selected objects and the underlying catalogue geometry. Environmental weighting changes clustering by design and can also alter the number density and effective selection; these must be controlled when matching the target sample. It is not a substitute for fully non-Gaussian
initial conditions, since it does not reproduce PNG-modified formation
histories, halo abundances, or all higher-order correlations. Within
the \texttt{HOLI} strategy, the two approaches are therefore
complementary: PNG initial conditions in \texttt{WebON} provide the
dynamical route, while large-scale web weighting provides a fast
emulator for massive mock production and controlled systematic tests.

We will extend these tests to evolving PNG lightcones, measuring the redshift dependence of the scale-dependent bias and its dependence on tracer selection. The programme will compare PNG initial conditions with redshift-dependent environmental weighting, test the expected $[T(k)D(z)]^{-1}$ scaling, and assess how broad redshift windows mix the evolving response. Matched Gaussian and non-Gaussian realisations will help isolate the response from sample variance. We will also investigate its separation from the GR position correction in Sec.~\ref{sec:webon:gr}; their large-scale enhancement is similar, but the two terms have different origins and must be compared using consistent gauge and potential conventions. These lightcone-evolution studies are planned extensions of the fixed-redshift validation in Fig.~\ref{fig:dalal_probe}.

\subsection{Accuracy considerations and practical choices}
\label{sec:webon:accuracy}

The main numerical checks concern snapshot interpolation, the
accuracy of emission-redshift assignment, and the finite simulation
volume.

\paragraph{Snapshot spacing and continuous-redshift correction.}
Equation~\eqref{eq:webon_continuous} assumes that the displacement difference between the target redshift $z_r$ and the stored displacement at $z_{\rm B}$ is well captured by a linear LPT increment.
This is satisfied when snapshots are spaced such that the correction remains a perturbative adjustment rather than the dominant contribution.
In practice, one can monitor the distribution of the relative correction amplitude, $\|\Delta\vPsi\|/\|\vPsi_{\rm app}\|$, excluding near-zero denominators or using an rms normalisation. Because the nonlinear residual is inherited from a stored field, the linear correction alone does not guarantee exact continuity at every bin boundary; convergence with snapshot spacing and direct boundary tests remain necessary.

\paragraph{Boundary conditions and observer placement.}
Because the lightcone is extracted from a periodic cube, the observer placement and maximum redshift determine which parts of the cube contribute.
 \texttt{WebON} avoids replication by construction. For a full-sky sphere centred in a cube of side length $L$, the selected radius must lie within the available volume, with an appropriate boundary margin; in the centred idealisation, $\chi_{\max}\leq L/2$. Sky maps and rotation tests can reveal boundary artefacts, while the absence of modes longer than the box remains a finite-volume limitation.

\paragraph{Recommended diagnostics.}
For production we recommend: (i) angular power spectra in several redshift shells, (ii) redshift evolution of $P(k)$ and redshift-space multipoles in thin bins, and (iii) cross-correlations between web levels and reference halo fields to ensure that tidal and curvature information is preserved across the working range of $k$.

\section{Hierarchical Cosmic-Web Modelling and the HOLI Strategy}
\label{sec:bias}
\label{sec:method}

The \texttt{HOLI} pipeline uses the phase-coherent matter field of
\texttt{WebON} to generate tracer populations through hierarchical
cosmic-web conditioning, field-level bias, and calibrated subgrid
models. {A schematic tracer-bias expansion illustrates the
local-in-matter-density and nonlocal contributions that motivate this
approach:}
\begin{equation}
{
\begin{aligned}
\delta_h(\vx) ={}&
\underbrace{b_1\delta+\frac{b_2}{2}\delta^2
+\frac{b_3}{6}\delta^3}_{\text{local density bias}}
\\ &+\underbrace{\frac{b_{s^2}}{2}s^2
+\frac{b_{\delta s^2}}{2}\delta s^2
+\frac{b_{s^3}}{6}s^3}_{\text{tidal bias}}
\\ &+\underbrace{\frac{b_{\mu^2}}{2}\mu^2
+\frac{b_{\delta\mu^2}}{2}\delta\mu^2
+\frac{b_{\mu^3}}{6}\mu^3}_{\text{density-curvature bias}}
+\cdots+\epsilon_h.
\end{aligned}}
\label{eq:bias_expansion}
\end{equation}
{Here $s^n\equiv\operatorname{Tr}(s^n)$ and
$\mu^n\equiv\operatorname{Tr}(\mu^n)$ are invariants of the
traceless tidal and density-curvature tensors defined below;
$\epsilon_h$ represents stochastic bias. The expansion is illustrative,
not a complete renormalised operator basis: mixed operators, further
higher derivatives, time-dependent contributions and renormalisation
subtractions are omitted
\citep{McDonald09,2018PhR...733....1D,Coloma_2024}.}

\subsection{Spectral hierarchy and its bias operators}
\label{sec:hierarchy}

For the matter overdensity $\delta$, define the spectral operator ladder
\begin{equation}
\widehat{\delta}^{(i)}(\vk) = -k^i\widehat{\delta}(\vk), \qquad i\in 2\mathbb{Z}.
\label{eq:spectral_ladder}
\end{equation}
The inverse operators are evaluated for $k>0$, with the zero mode set to zero. A common smoothing and mesh-assignment convention must be used when comparing levels, especially because positive powers of $k$ amplify fluctuations near the mesh cutoff. Each level emphasises a different spatial regime:
\begin{itemize}
\item $i=-2$: long-range tidal fields (potential/tidal-tensor web),
\item $i=0$: curvature (density-Hessian web),
\item $i=2,4$: higher-derivative short-range response levels.
\end{itemize}

{At each level, the symmetric Hessian}
\begin{equation}
\mathcal{H}^{(i)}_{ab} = \partial_a\partial_b\delta^{(i)},
\label{eq:hessian_level}
\end{equation}
{is diagonalised to classify each cell as a knot, filament,
sheet or void according to whether three, two, one or zero eigenvalues
exceed a specified threshold. The associated eigenvalues also determine
the scalar tensor invariants relevant to nonlocal tracer bias.}

{At $i=-2$, $\delta^{(-2)}=\nabla^{-2}\delta\equiv\Phi$
with $\nabla^2\Phi=\delta$, so $\mathcal{H}^{(-2)}_{ab}$ is
the tidal tensor $T_{ab}=\partial_a\partial_b\Phi$. Its traceless
part is}
\begin{equation}
{s_{ab}=T_{ab}-\frac{\delta}{3}\delta_{ab}^{\rm K},}
\end{equation}
{whose quadratic and cubic invariants are
$s^2=s_{ab}s_{ba}$ and $s^3=s_{ab}s_{bc}s_{ca}$.
At $i=0$, $\delta^{(0)}=-\delta$, so the density-curvature
tensor $\Gamma_{ab}=\partial_a\partial_b\delta$
satisfies $\mathcal{H}^{(0)}_{ab}=-\Gamma_{ab}$. Its
traceless part,}
\begin{equation}
{\mu_{ab}=\Gamma_{ab}
-\frac{\nabla^2\delta}{3}\delta_{ab}^{\rm K},}
\end{equation}
{defines $\mu^2=\mu_{ab}\mu_{ba}$ and
$\mu^3=\mu_{ab}\mu_{bc}\mu_{ca}$. The opposite sign of
$\mathcal{H}^{(0)}$ relative to $\Gamma$ must be respected
when assigning curvature-web labels. These tensors connect the
long-range tidal and shorter-range density-curvature information to
the corresponding invariants in Eq.~\eqref{eq:bias_expansion}
\citep{Coloma_2024}.}

\begin{figure}[tbp]
\centering
\includegraphics[width=\textwidth]{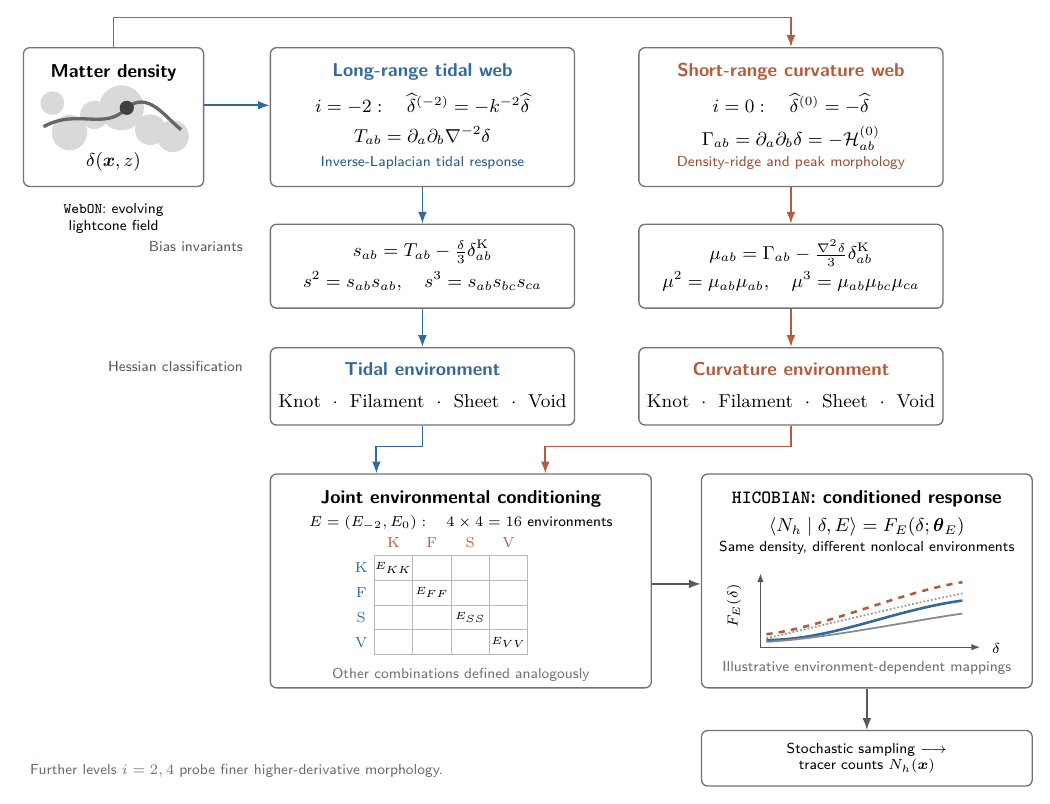}
\caption{{
\textbf{Spectral cosmic-web classification as a compact representation
of nonlinear, nonlocal tracer bias.}
The evolving \texttt{WebON} matter-density field
$\delta(\boldsymbol{x},z)$ feeds two parallel spectral branches:
the inverse-Laplacian tidal level ($i=-2$, blue) and the
density-curvature level ($i=0$, orange). The corresponding traceless
tensors yield the quadratic and cubic bias invariants
$s^2,s^3$ and $\mu^2,\mu^3$, respectively.
Eigenvalue-threshold classification independently identifies knots (K),
filaments (F), sheets (S) and voids (V) at each level. The unfilled
$4\times4$ grid represents the 16 joint environments
$E=(E_{-2},E_0)$, with examples $E_{KK}$, $E_{FF}$, $E_{SS}$ and
$E_{VV}$; all other class pairs are included. Within each environment,
\texttt{HICOBIAN} models the conditional mean tracer count
$\langle N_h\mid\delta,E\rangle=F_E(\delta;\boldsymbol\theta_E)$.
The distinct plotted curves are illustrative, not fitted responses.
Stochastic sampling converts the conditional means into discrete tracer
counts $N_h(\boldsymbol{x})$. Levels $i=2,4$ extend the hierarchy
towards finer higher-derivative morphology.}}
\label{fig:web_nonlocal_bias}
\end{figure}

{Figure~\ref{fig:web_nonlocal_bias} illustrates this
connection. Both spectral levels are derived independently from the
same matter field, rather than by applying the curvature operator to
the tidal web. Each yields four eigenvalue-defined environments;
combining their classifications produces $4\times4=16$ joint
subregions $E=(E_{-2},E_0)$. Within each subregion,
\texttt{HICOBIAN} calibrates a positive-definite conditional
mean tracer count $F_E(\delta;\boldsymbol\theta_E)$,
which is converted into discrete counts by stochastic sampling.
Thus, the response is local in $\delta$ conditional on $E$, but
implicitly nonlocal in the underlying matter field. The complete
eigenvalue triplets determine the scalar invariants, whereas
four-valued web labels retain only a compressed representation;
the predictive accuracy must therefore be established by
calibration and validation.}

{With the sign convention of Eq.~\eqref{eq:spectral_ladder},
eigenvalue thresholds must be defined consistently at each level,
whose tensors also have different dimensions. The boundaries between
environments need not be sharp: fuzzy membership functions can assign
continuous degrees of membership to each web class
\citep[][]{2026A&A...709A.110R}. Thresholds may be fixed or calibrated
in redshift shells, provided their evolution does not induce artificial
changes in web-volume fractions. The hierarchy is motivated by
perturbative operator structure but its calibrated nonlinear responses
are not restricted to a finite-order bias expansion.}

{The $i=-2$ and $i=0$ morphologies are illustrated separately
in Figs.~\ref{fig:tweb_minus2} and~\ref{fig:tweb_zero}, while
$i=2,4$ emphasise progressively finer spatial structure.}

\begin{figure}[tbp]
\centering
\includegraphics[width=\textwidth]{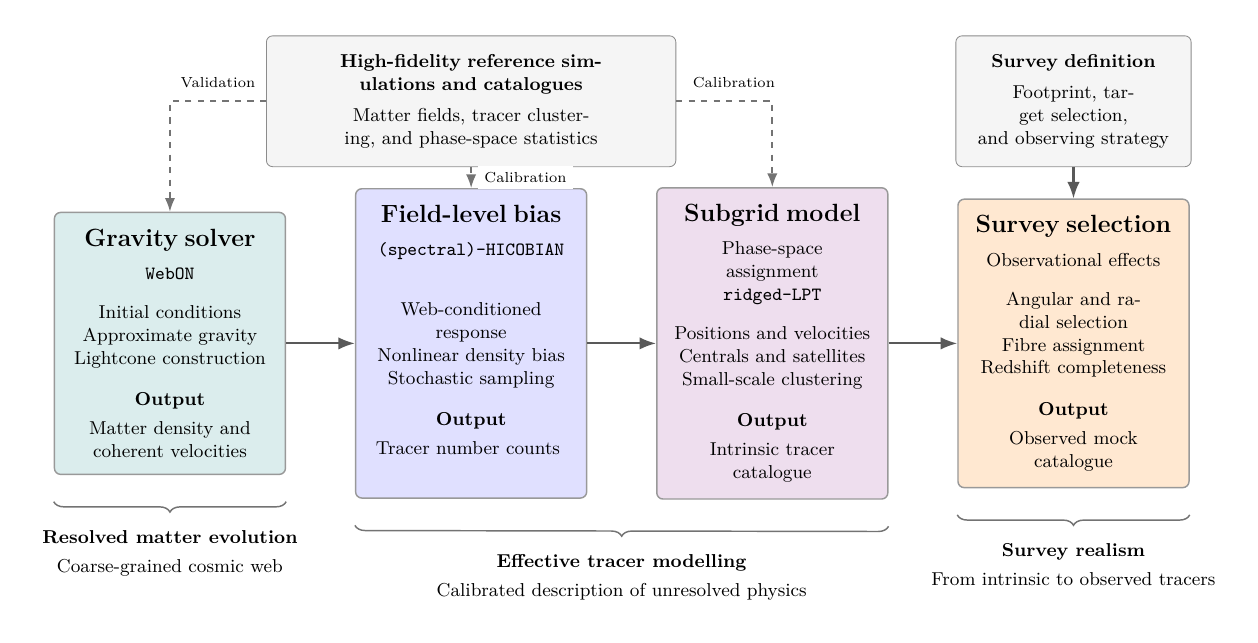}
\caption{Methodological structure of the HOLI mock framework. Solid arrows indicate the production workflow. \texttt{WebON} constructs the evolving, coarse-grained matter lightcone and its coherent velocity field. The spectral cosmic-web hierarchy supplies environmental information to \texttt{HICOBIAN}, which generates tracer number counts through an effective field-level bias model and stochastic sampling. Subgrid prescriptions assign tracer positions and velocities, including central and satellite populations where required. Survey selection then maps the intrinsic catalogue to an observed mock catalogue. Dashed arrows indicate validation of the matter backbone and calibration of the bias and subgrid components against high-fidelity references. The framework separates resolved matter evolution from the effective treatment of unresolved tracer physics and the application of observational selection.}
\label{fig:holi_workflow}
\end{figure}

\subsection{Mock-making context and HOLI workflow}
\label{sec:method:pipeline}

Figure~\ref{fig:holi_workflow} summarises the four stages:
(1) \texttt{WebON} matter and velocity lightcones;
(2) environmental segmentation and effective tracer number-count
modelling; (3) subgrid tracer positions and velocities; and
(4) optional tracer-property assignment followed by survey
selection. The matter mesh (typically $\sim1$--$6\,h^{-1}\mathrm{Mpc}$)
is not required to resolve each galaxy-host halo; the distinction
between resolved matter dynamics and subgrid tracer physics is
central to the method. Some DESI tracer populations require
modelling haloes down to masses of order
$10^{11}\,h^{-1}\,\mathrm{M}_\odot$, making this separation
particularly valuable for large mock ensembles.

\paragraph{Relation to existing mock methods.}
Particle-mesh (PM) and PM-like gravity calculations paired with
halo-occupation recipes include COLA-HOD \citep{Izard_2016},
GLAM-HOD \citep{Ereza2024}, and FastPM-HOD \citep{Variu_2024}.
Some use replicated boxes for survey-size lightcones; large-volume
realisations can also avoid replication, but at a cost set by the
chosen volume and mass/force resolution.
Methods based on resolved
subhaloes or high-resolution phase space, such as SHAMe
\citep{2021MNRAS.504.5205C,2023MNRAS.525.3149C,2024A&A...689A..66O}
and HOMe \citep{2026ApJ..1002...90F}, target information that a
coarse mesh cannot supply without further modelling. Alternative
approximate halo-identification schemes include PeakPATCH
\citep{Bond_1996nat,Bond_1996b,Bond_1996c}, PINOCCHIO
\citep{Monaco_2002,Monaco_2013}, halo-finding PThalos
\citep{Manera_2013,Manera_2015}, and HALOGEN
\citep{Avila:2014nia}; resolution and high-$k$ performance remain
important considerations \citep{Chuang_2015,Blot_2019,Colavincenzo_2019,Lippich_2019}.

\texttt{HOLI} instead follows field-level approaches including
PThalos \citep{Scoccimarro_2002}, PATCHY
\citep{Kitaura_2014,Kitaura_2015,Kitaura_2016}, EZmocks
\citep{Chuang_2015,Zhao_2021}, BAM
\citep{Balaguera_2018,Balaguera_2020,Balaguera_2023,Kitaura_2022},
and \texttt{HICOBIAN}
\citep{Coloma_2024,Sinigaglia_2024,2026JCAP...06..037S}.
Here the gravity solver supplies the resolved density and
nonlocal environment fields, while the tracer response is
calibrated using reference catalogues. Such calibration allows
fine-scale tracer sampling but cannot independently recover
unresolved physical information. The framework builds on the
COSMIC SIGNAL programme\footnote{COSMIC SImulated Galaxy Networks
Applied on Lightcones; \url{www.cosmic-signal.org}} and the earlier
\texttt{COSMIC-BIRTH} lightcone approach \citep{Kitaura_2021}.

\paragraph{Hierarchical field-level bias.}
\label{sec:method:bias}
\label{sec:hicobian}
\texttt{HICOBIAN} \citep{Coloma_2024,Sinigaglia_2024}
{uses the environmental classification described in
Sec.~\ref{sec:hierarchy}. Within each subset,} a positive-definite
density-response model (often with up to approximately six parameters)
predicts expected tracer number counts, which a calibrated stochastic
prescription then converts to a discrete catalogue. Parameters may be inferred by
MCMC \citep{Vakili_2017} or genetic optimisation
\citep{Coloma_2024,Sinigaglia_2024}. {Its stochasticity and higher-order statistics must be
validated against reference catalogues.} Importantly, this effective tracer model does not require the raw
ALPT matter power spectrum to reproduce the $N$-body result at the
same small-scale precision. Coloma-Nadal et al.~2024 \citep{Coloma_2024} demonstrated
better than $2\%$ agreement for the power spectrum of calibrated
halo catalogues up to $k\simeq0.8\,h\,\mathrm{Mpc}^{-1}$ using
coarse ALPT fields (with a mesh cell size of approximately
$3.9\,h^{-1}\,\mathrm{Mpc}$) and a hierarchical cosmic-web
classification, compared with a reference halo sample from the
UNIT simulation. The precise accuracy and limiting wavenumber
depend on the tracer, calibration sample, redshift, and summary
statistic; this result should not be read as a universal guarantee
for every tracer. Post-processing ridging provides a complementary
way to improve short-range matter morphology and tracer positioning
\citep{2026arXiv260313106K}. Further tracer-dependent calibration
and subgrid applications are developed in companion studies
(Forero-S\'anchez et al. and Tang et al., in preparation), including
tests in the approximate range
$k\simeq0.4$--$0.8\,h\,\mathrm{Mpc}^{-1}$.

\paragraph{Subgrid positions and velocities.}
\label{sec:method:subgrid}
Field-level number counts are mapped to discrete positions using
the particle distribution and interpolated displacements
\citep{Forero_2024}. Short-range clustering is corrected either
through local attractors and relocation, as in \texttt{CosmoMIA}
\citep{Forero_2024}, or via calibrated ridging
\citep{2026arXiv260313106K}. Tracer peculiar velocities combine
coherent (e)ALPT flows with density- and environment-dependent
stochastic components constrained by redshift-space multipoles.
Satellite prescriptions (e.g. NFW/HOD) can be added when required;
none of these steps substitutes for resolved subhalo dynamics.
Their tracer-dependent validation is discussed in companion work.

\paragraph{Tracer properties and observational selection.}
\label{sec:method:properties}
When needed, \texttt{HOLI} assigns properties such as luminosity,
stellar mass, and halo-mass proxies using environment-dependent
models, including HADRON \citep{Zhao_2015,Balaguera_2023} or
alternative prescriptions \citep{Ramakrishnan_2021}.
Clustering after property selection must be validated separately
from the one-point property distribution. Finally, survey geometry,
radial selection, masks, fibre assignment, redshift failures, and
other observational effects are applied as survey-specific steps.

\subsection{Calibration, applicability, and limitations}
\label{sec:method:calibration}
\label{sec:method:limitations}

The pipeline is applied in three stages: (i) calibrate bias,
stochasticity, subgrid positions, velocities, and properties in
periodic boxes at selected redshifts using high-fidelity reference
catalogues; (ii) test power spectra, multipoles, higher-order
statistics, and configuration-space clustering; and (iii) deploy
the calibrated models onto \texttt{WebON} lightcones with
interpolated redshift-dependent parameters and survey selection.
Separate tests of the lightcone backbone and the tracer model help
to identify the source of discrepancies. For example, failures in
coherent flows or cross-correlations point towards the gravity
backbone, whereas incorrect small-scale redshift-space multipoles
may indicate subgrid-velocity or relocation prescriptions.

The resolved quantity is the Mpc-scale, phase-coherent dark-matter
field, not the detailed internal structure of individual haloes.
The mesh cutoff limits directly predicted small-scale information;
halo exclusion, close pairs, satellite offsets, virial motions,
and shot-noise or non-Poisson stochasticity must be modelled
phenomenologically and checked against references. The framework
is appropriate for large-volume clustering statistics but does not
by itself predict merger trees, internal profiles, baryonic
feedback, or detailed satellite dynamics. Calibration outside the
reference cosmology, tracer selection, or mesh resolution requires
new tests. Snapshot spacing, observer placement, and lightcone
crossing accuracy are discussed in Sec.~\ref{sec:webon:accuracy}.

\section{Numerical Consistency Tests of the \texttt{WebON} Lightcones}
\label{sec:validation}

We test the internal consistency of the matter-lightcone construction
and its cosmic-web representation. The absolute accuracy of the
underlying approximate matter dynamics is a distinct question,
already assessed for post-processed ridged LPT fields against
high-fidelity Abacus $N$-body simulations
\citep{2026arXiv260313106K}; the accuracy of effective halo
clustering from coarse ALPT fields has likewise been quantified
for \texttt{HICOBIAN} \citep{Coloma_2024}. These external
benchmarks are complementary to the present lightcone tests, and
should not be conflated with them.

\subsection{Matter and cosmic-web evolution}
\label{sec:validation:morphology}

\begin{figure}[tbp]
  \centering
  \vspace{-0.5cm}
  \begin{tabular}{c}
    \includegraphics[width=0.9\textwidth]{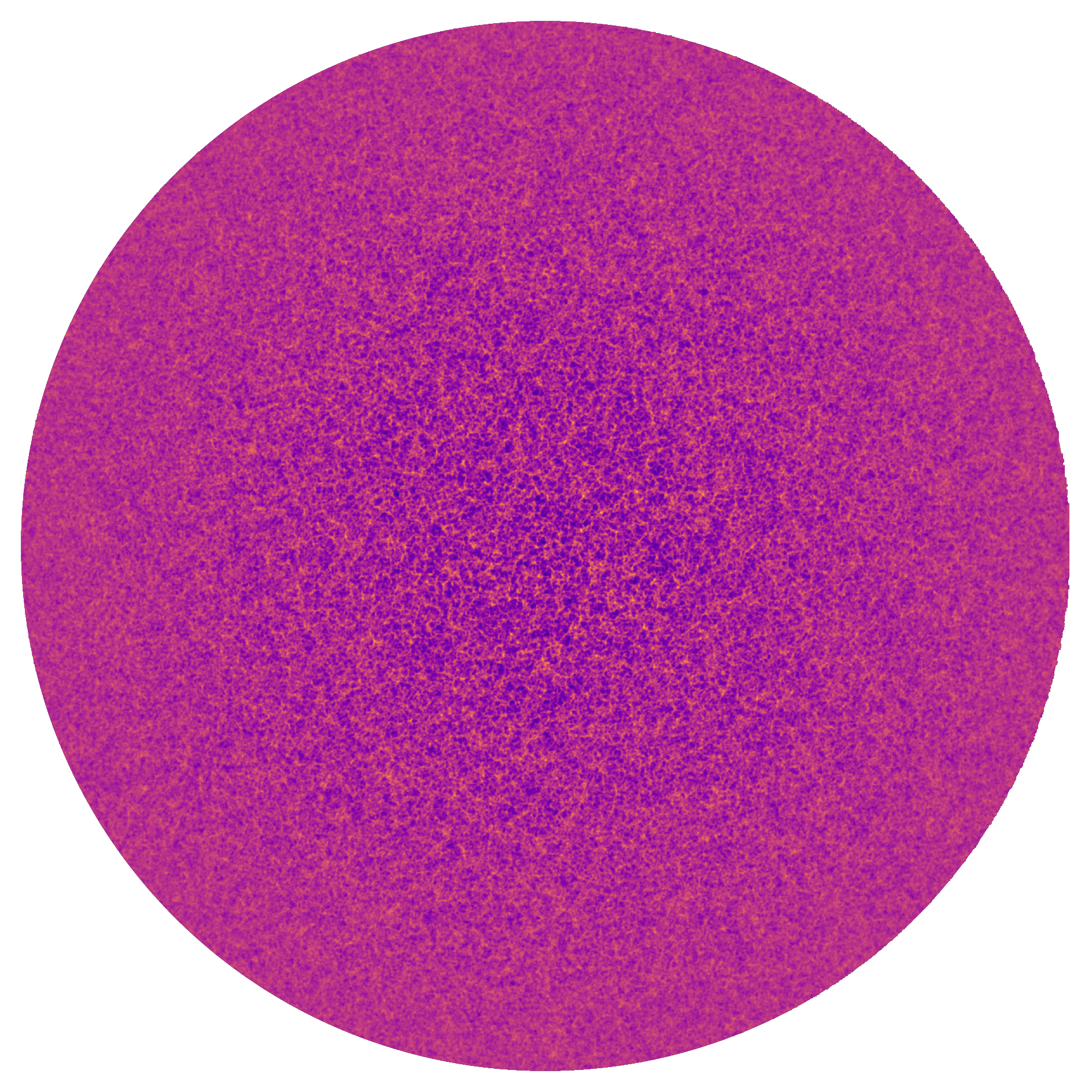}
    \put(-420,364){\rotatebox[]{90}{\Huge$\longrightarrow$}}
    \put(-425,205){\rotatebox[]{90}{\Large$10\,h^{-1}\,{\rm Gpc}$}}
    \put(-420,21){\rotatebox[]{90}{\Huge$\longleftarrow$}}
    \put(-400,370){\rotatebox[]{0}{\Large$\delta_{\rm ALPT}$}}
    \vspace{-0.3cm}
    \\\includegraphics[width=0.575\textwidth]{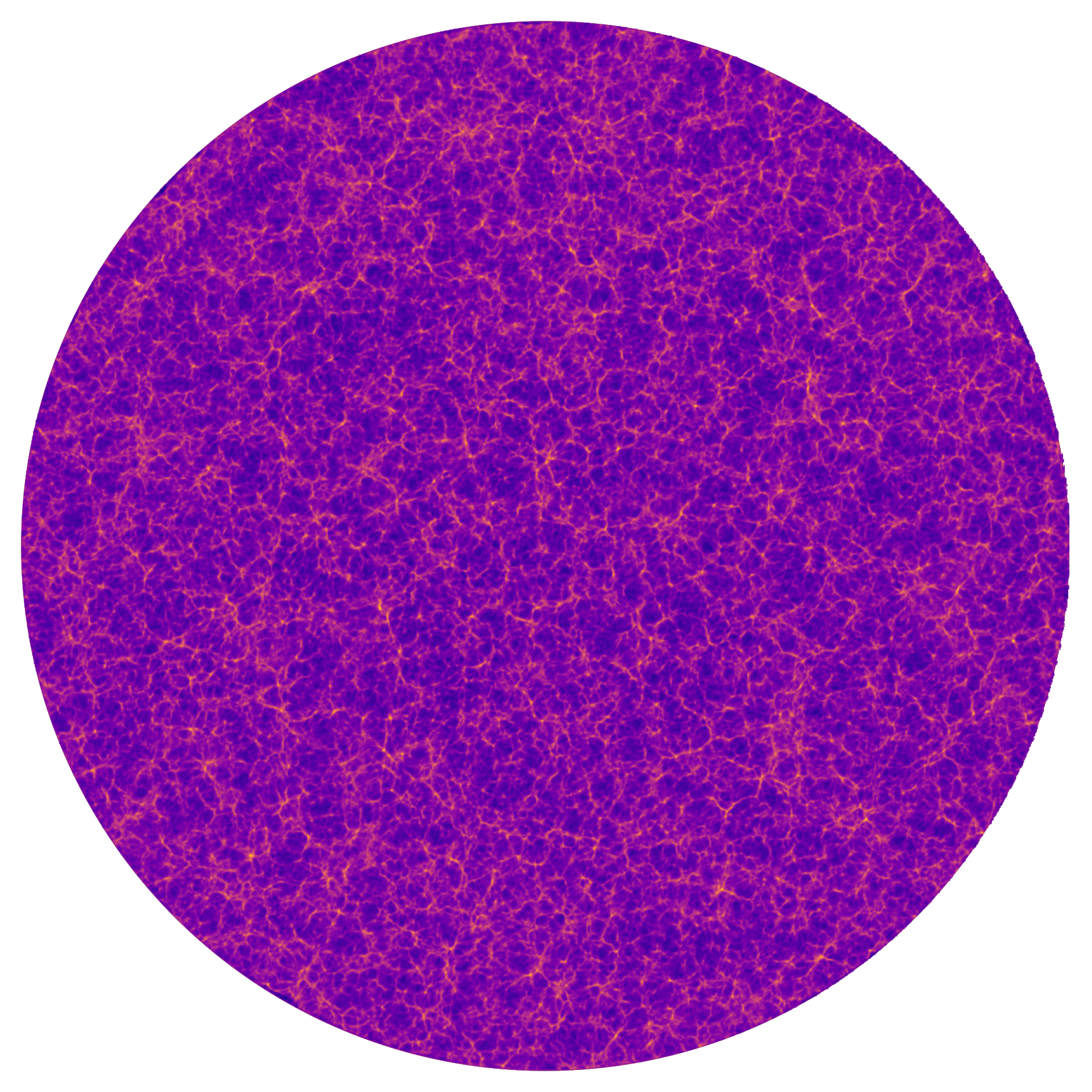}
    \put(-280,220){\rotatebox[]{90}{\Huge$\longrightarrow$}}
    \put(-285,123){\rotatebox[]{90}{\Large$2.2\,h^{-1}\,{\rm Gpc}$}}
    \put(-280,22){\rotatebox[]{90}{\Huge$\longleftarrow$}}
    \put(-260,220){\rotatebox[]{0}{\Large$\delta_{\rm eALPT}$}}
  \end{tabular}
  \vspace{-0.6cm}
  \caption{Slices through  periodic volumes
containing full-sky \texttt{WebON} lightcones with the observer at the
centre, illustrating the construction described in Sec.~\ref{sec:webon}. \emph{Top:} ALPT lightcone in a volume $V=(10\,h^{-1}\,\mathrm{Gpc})^3$, with cell size
$5.56\,h^{-1}~{\rm Mpc}$, as used for DESI mock generation. \emph{Bottom:} eALPT lightcone in a volume $V=(2.2\,h^{-1}\,{\rm Gpc})^3$, with cell size $\sim1.85\,h^{-1}\,{\rm Mpc}$. .}
  \label{fig:lightcone_maps}
\end{figure}

Figure~\ref{fig:lightcone_maps} shows representative slices through
the two \texttt{WebON} configurations considered in this work. The
large-volume configuration demonstrates the ability to construct a
DESI-scale full-sky lightcone without box replication, whereas the
higher-resolution configuration resolves finer filamentary and knot
structure.

The continuous-redshift correction in
Eq.~\eqref{eq:webon_continuous} is designed to prevent discontinuities
at the boundaries of the stored displacement intervals. We test this
by inspecting the evolution of the density field and its angular
statistics in consecutive redshift shells. The angular-clustering comparison in Fig.~\ref{fig:angular_power_spectrum} complements these visual checks by testing the projected clustering amplitude in consecutive shells. Direct radial boundary checks and snapshot-spacing convergence provide additional tests of continuity.

\paragraph{Evolution of the spectral cosmic web.}
\label{sec:validation:web}

\begin{figure}[tbp]
  \centering
  \vspace{-0.5cm}
  \begin{tabular}{c}
    \includegraphics[width=0.9\textwidth]{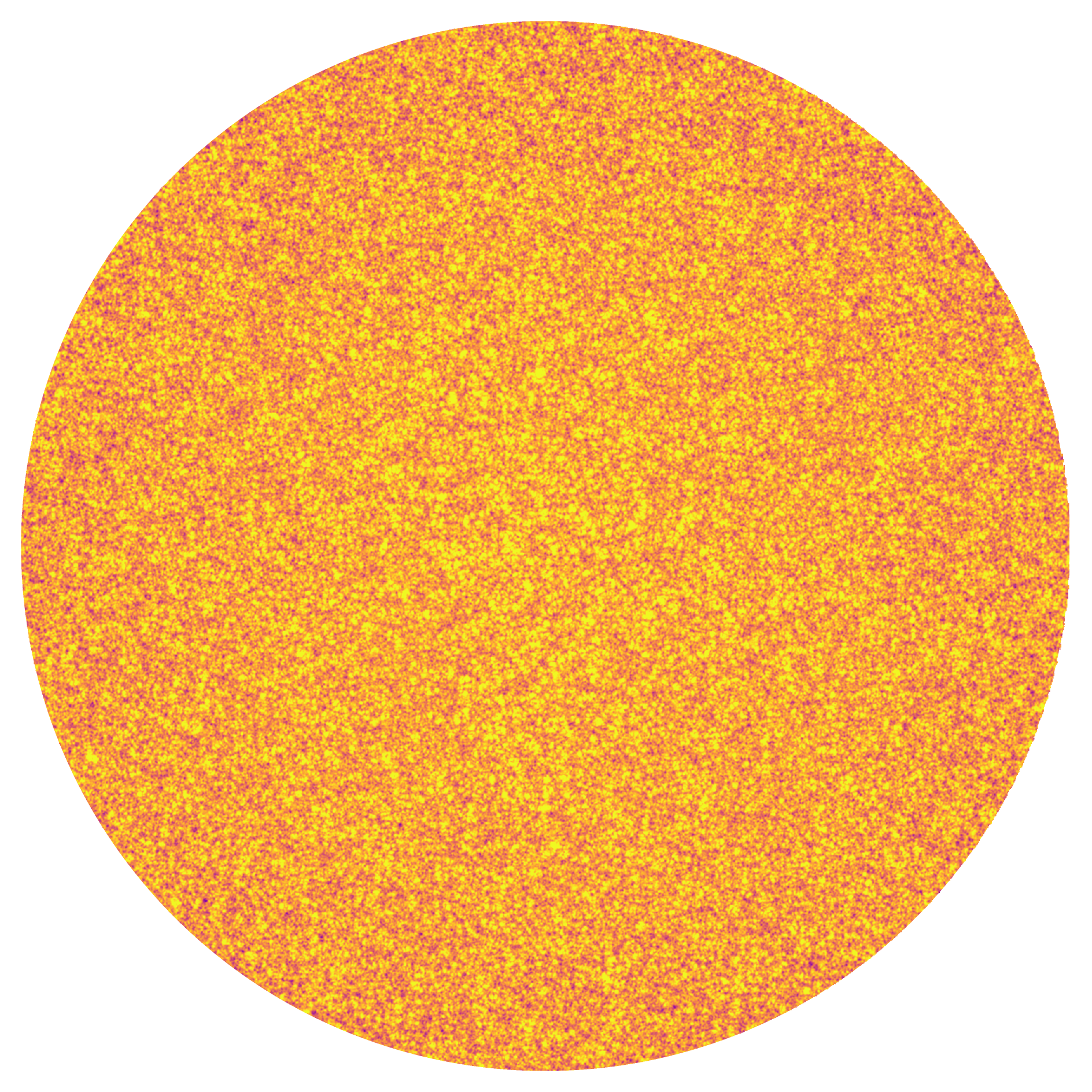}
    \put(-420,364){\rotatebox[]{90}{\Huge$\longrightarrow$}}
    \put(-425,205){\rotatebox[]{90}{\Large$10\,h^{-1}\,{\rm Gpc}$}}
    \put(-420,21){\rotatebox[]{90}{\Huge$\longleftarrow$}}
    \put(-400,370){\rotatebox[]{0}{\Large$-k^{-2}$}}
    \vspace{-0.3cm}
    \\\includegraphics[width=0.575\textwidth]{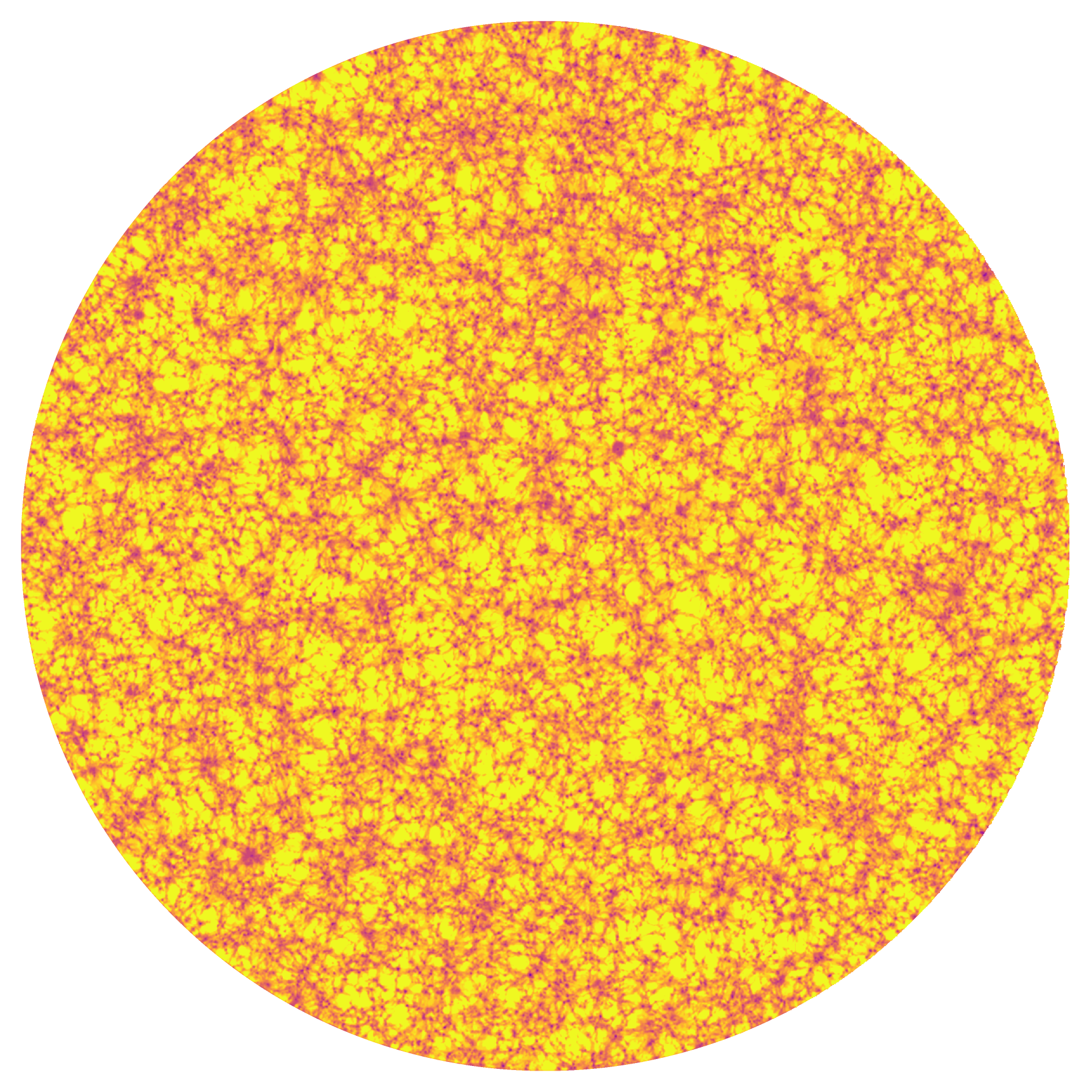}
    \put(-280,220){\rotatebox[]{90}{\Huge$\longrightarrow$}}
    \put(-285,123){\rotatebox[]{90}{\Large$2.2\,h^{-1}\,{\rm Gpc}$}}
    \put(-260,220){\rotatebox[]{0}{\Large$-k^{-2}$}}
    \put(-280,22){\rotatebox[]{90}{\Huge$\longleftarrow$}}
  \end{tabular}
  \vspace{-0.3cm}
  \caption{Slices through the potential-like field underlying the tidal-web classification ($i=-2$ level of the spectral hierarchy) computed on the two representative lightcone boxes shown in Fig.~\ref{fig:lightcone_maps}, in the same top/bottom order. The inverse-Laplacian weighting emphasises long-range structure; taking the Hessian of this scalar gives the tidal tensor used for environmental classification. See Eqs.~\eqref{eq:spectral_ladder} and \eqref{eq:hessian_level}.}
  \label{fig:tweb_minus2}
\end{figure}

\begin{figure}[tbp]
  \centering
  \vspace{-0.5cm}
  \begin{tabular}{c}
    \includegraphics[width=0.9\textwidth]{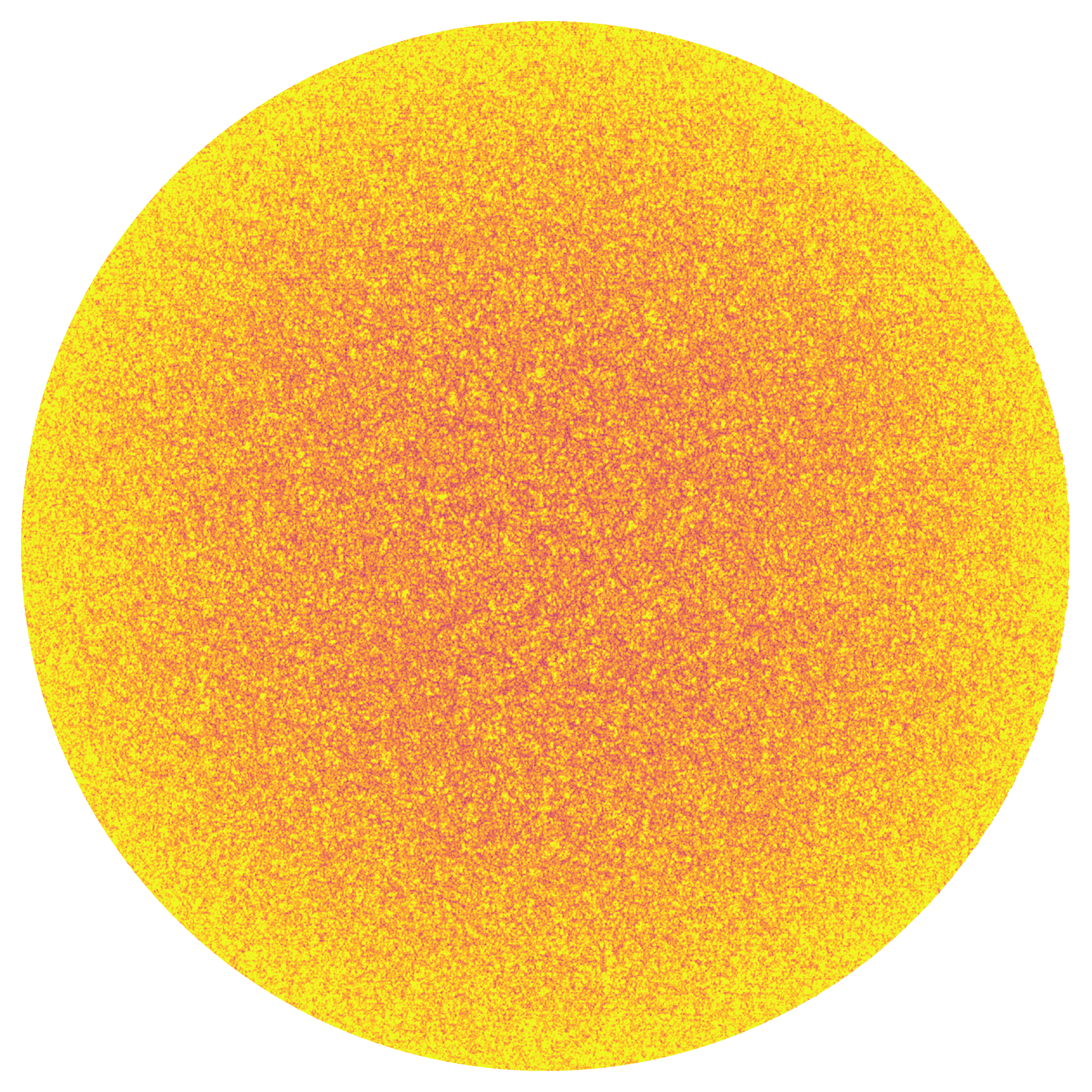}
    \put(-420,364){\rotatebox[]{90}{\Huge$\longrightarrow$}}
    \put(-425,205){\rotatebox[]{90}{\Large$10\,h^{-1}\,{\rm Gpc}$}}
    \put(-420,21){\rotatebox[]{90}{\Huge$\longleftarrow$}}
    \put(-400,370){\rotatebox[]{0}{\Large$-k^{0}$}}
    \vspace{-0.3cm}
    \\\includegraphics[width=0.575\textwidth]{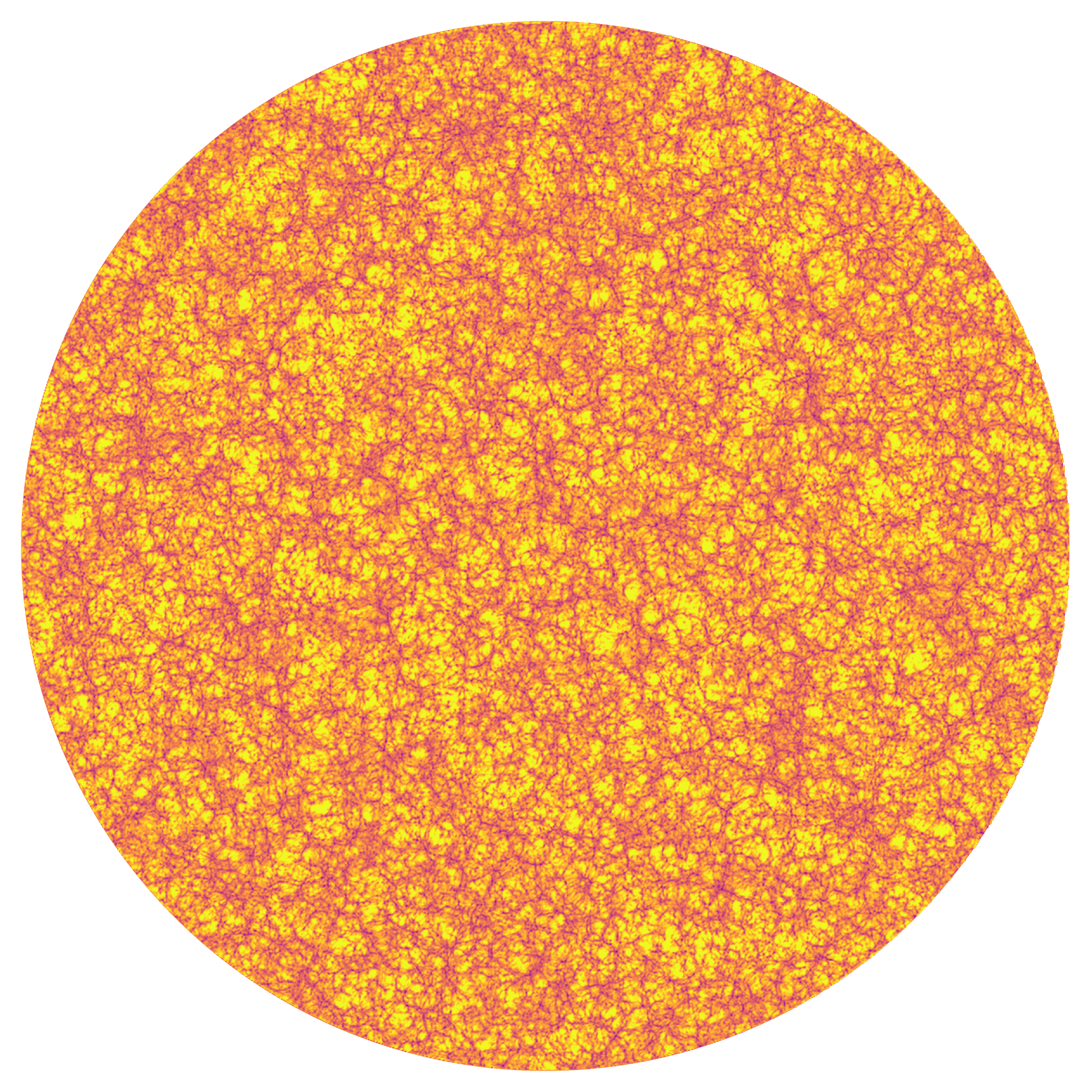}
    \put(-280,220){\rotatebox[]{90}{\Huge$\longrightarrow$}}
    \put(-285,123){\rotatebox[]{90}{\Large$2.2\,h^{-1}\,{\rm Gpc}$}}
    \put(-260,220){\rotatebox[]{0}{\Large$-k^{0}$}}
    \put(-280,22){\rotatebox[]{90}{\Huge$\longleftarrow$}}
  \end{tabular}
  \vspace{-0.6cm}
  \caption{Slices through the field underlying the density-curvature web ($i=0$ level of the spectral hierarchy), computed on the same two lightcone boxes and displayed in the same order as Fig.~\ref{fig:lightcone_maps}. At this level the scalar kernel is $-k^0$, so the curvature information enters through the Hessian of the resulting field. Compared with the potential-like field in Fig.~\ref{fig:tweb_minus2}, this level retains more local density structure and supplies complementary information for environment-conditioned bias modelling.}
  \label{fig:tweb_zero}
\end{figure}

Figures~\ref{fig:tweb_minus2} and \ref{fig:tweb_zero} show the
potential/tidal and curvature levels of the spectral hierarchy,
respectively. The same lightcone density realisation is used at each
level, allowing the long-range tidal morphology and the more local
ridge and curvature structure to be followed continuously across
redshift.

These maps illustrate the phase-coherent construction of the evolving cosmic web from a common realisation, rather than from independently generated fields in disconnected redshift shells. Their visual continuity is a morphological diagnostic; a numerical accuracy claim additionally requires comparison with reference fields. Quantitative
tests of the usefulness of these fields for tracer bias are presented
in the companion \texttt{HOLI}/\texttt{HICOBIAN} analyses.

\subsection{Angular clustering}
\label{sec:validation:angular}

\begin{figure}[tbp]
  \centering
  \vspace{-1cm}
  \begin{tabular}{c}
    \includegraphics[width=0.75\textwidth]{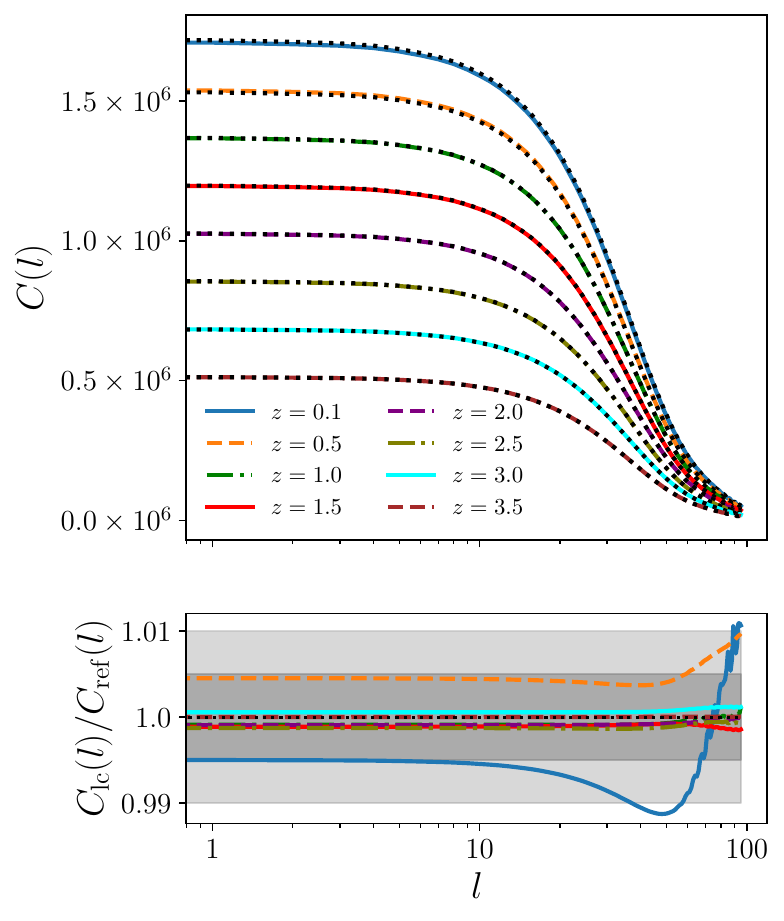}
  \end{tabular}
  \caption{Internal angular-clustering consistency test for
\texttt{WebON}. The plotted ratios compare the angular power spectra
$C_\ell$ of continuously evolving \texttt{WebON} lightcone shells with
those of corresponding shells extracted from fixed-redshift
\texttt{WebON-JAX} boxes. Both use the same ALPT gravity approximation,
observer position, and radial selection. The shell centres are
$z=0.1,\,0.5,\,1.0,\,1.5,\,2.0,\,2.5,\,3.0,$ and $3.5$.
In the upper panel, coloured solid (or coloured dashed/dash-dotted)
curves denote the evolving lightcone spectra at the labelled
redshifts, while black dotted curves denote their respective
fixed-redshift references. The lower panel shows the corresponding
ratio $C_{\ell,\mathrm{lc}}/C_{\ell,\mathrm{ref}}$;
the dark- and light-grey bands mark $\pm0.5\%$ and
$\pm1\%$ about unity, respectively.
The plotted ratios show percent-level consistency, with a few
multipoles approaching or marginally exceeding the $\pm1\%$ band. This establishes internal lightcone consistency; external
matter and tracer accuracy benchmarks are given in
Kitaura \& Sinigaglia 2026 \citep{2026arXiv260313106K} and Coloma-Nadal et al.~2024 \citep{Coloma_2024}, respectively.
}
\label{fig:angular_power_spectrum}
\end{figure}

We quantify the internal consistency of lightcone evolution using
angular power spectra measured in redshift shells.
Figure~\ref{fig:angular_power_spectrum} compares
the \texttt{WebON} lightcone measurements with those obtained from
fixed-redshift boxes computed with \texttt{WebON-JAX}, using the same
ALPT approximation. We consider snapshots at
$z=0.1,\,0.5,\,1.0,\,1.5,\,2.0,\,2.5,\,3.0,$ and $3.5$.
For each snapshot redshift, we compare the angular clustering in a
shell centred on that redshift within the lightcone volume with the
corresponding shell extracted from the fixed-redshift box, using the
same observer position and radial selection. This comparison tests
whether the continuously evolving lightcone reproduces the clustering
of the snapshot calculation at the corresponding epoch, subject to
the evolution across the finite redshift width of each shell.

Across the redshift and multipole ranges shown, the measured angular
clustering agrees with the fixed-redshift calculation at the percent level.
Since both calculations use the same ALPT gravity approximation, this
is an internal consistency test of the lightcone construction, not an
independent validation of the absolute matter-clustering amplitude or
shape. The comparison is nevertheless important for the tracer-bias
strategy: parameters calibrated at fixed-redshift snapshots can be
interpolated and applied to the corresponding evolving ALPT lightcone
fields, which may subsequently be processed with ridging and
hierarchical web-conditioned bias. The latter operations can improve
small-scale matter morphology and reproduce calibrated tracer
clustering beyond the mesh resolution, respectively; they do not
retroactively validate the raw ALPT matter spectrum. Independent
accuracy tests already exist: the companion ridged-LPT study compares
matter power spectra and field-level cross-correlations with Abacus
$N$-body simulations \citep{2026arXiv260313106K}, while Coloma-Nadal et al.~2024
\citep{Coloma_2024} demonstrate better than $2\%$ agreement in
the power spectrum of calibrated halo samples up to
$k\simeq0.8\,h\,\mathrm{Mpc}^{-1}$ using coarse ALPT fields and
\texttt{HICOBIAN}. Ongoing companion analyses extend this
calibration to additional tracers and subgrid prescriptions
(Forero-S\'anchez et al. and Tang et al., in preparation). Thus,
Fig.~\ref{fig:angular_power_spectrum} addresses the specific
remaining question of consistency between lightcone and snapshot
implementations rather than duplicating those external benchmarks.

\subsection{General-relativistic position correction}
\label{sec:validation:gr}

\begin{figure}[tbp]
  \centering
  \begin{tabular}{c}
    \includegraphics[width=0.75\textwidth]{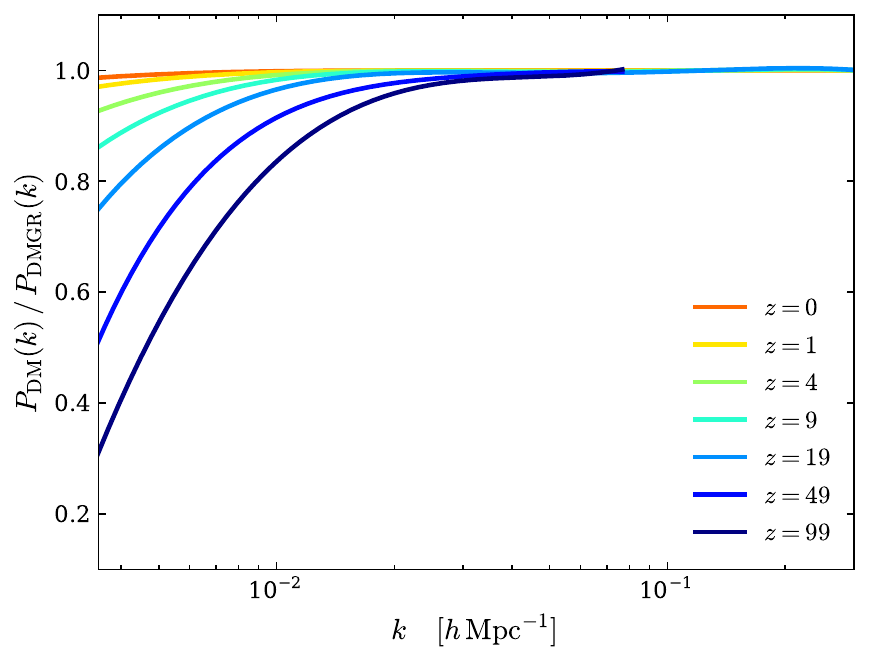}
  \end{tabular}
\caption{
Power-spectrum response to the Hahn--Paranjape relativistic position correction implemented in \texttt{WebON}.
We show the ratio $P_{\rm DM}(k)/P_{\rm DM}^{\rm GR}(k)$ for dark-matter fields evolved with the standard Newtonian displacement and with the additional GR screening displacement described in Sec.~\ref{sec:webon:gr}.
The correction is negligible on quasi-linear and nonlinear scales, where the ratio approaches unity, but becomes visible on the largest modes because the displacement correction is infrared-enhanced.
The curves show $z=0,1,4,9,19,49$, and $99$. The increasingly pronounced departure below unity at high redshift corresponds to enhanced power in the corrected density field, consistent with the linear prediction $[1+\ell^{-2}(z)/k^2]^{-2}$ derived in Sec.~\ref{sec:validation:gr}. The approach to unity towards higher $k$ illustrates the decreasing importance of the correction on subhorizon scales. This is a test of the position transformation, rather than a screened-force evolution or a complete relativistic lightcone calculation.
}
\label{fig:gr_probe}
\end{figure}

Figure~\ref{fig:gr_probe} shows the ratio of the uncorrected matter power spectrum to that obtained after applying the position correction in Eq.~\eqref{eq:webon_gr_fourier}. The curves span $z=0,1,4,9,19,49$, and $99$: the highest redshifts are controlled tests of the operator, rather than the redshift range of the DESI lightcones. The ratio falls below unity at low $k$ and approaches unity towards higher $k$, with a larger departure at higher redshift.

The sign follows directly from the displacement convention. At linear order, $\delta=-\nabla\cdot\vPsi$, so the added displacement gives
\begin{equation}
\widehat\delta_{\rm N}(\vk,z)
\simeq\left[1+\frac{\ell^{-2}(z)}{k^2}\right]
\widehat\delta_{\rm sim}(\vk,z),
\label{eq:gr_density_response}
\end{equation}
and hence the expected large-scale power-spectrum ratio is
\begin{equation}
\frac{P_{\rm DM}(k,z)}{P_{\rm DM}^{\rm GR}(k,z)}
\simeq\left[1+\frac{\ell^{-2}(z)}{k^2}\right]^{-2}.
\label{eq:gr_power_response}
\end{equation}
Thus, the ratio below unity corresponds to an enhancement of the corrected density power, not a suppression caused by replacing the force kernel. The plotted trends are consistent with this prediction. A quantitative verification of the amplitude requires comparison with Eq.~\eqref{eq:gr_power_response} for the same cosmology and redshifts; the figure alone does not establish a numerical error tolerance. This test concerns the approximate coordinate transformation, rather than a complete relativistic observable on the lightcone.

\subsection{Local primordial non-Gaussianity}
\label{sec:validation:png}

We test whether our framework recovers the characteristic
large-scale bias induced by local primordial non-Gaussianity
\cite{Dalal_2008}. In this model, coupling between long-wavelength
potential fluctuations and short-wavelength density fluctuations
produces a scale-dependent tracer response,
\begin{equation}
    \Delta b(k,z)
    \propto \frac{f_{\rm NL}}{k^2 T(k)D(z)},
    \label{eq:png_bias_scaling}
\end{equation}
where $T(k)$ is the matter transfer function and $D(z)$ is the
linear growth factor. Since $T(k)\rightarrow1$ as $k\rightarrow0$,
the correction approaches the characteristic $k^{-2}$ dependence.

\begin{figure}[tbp]
  \centering
  \begin{tabular}{c}
    \includegraphics[width=0.75\textwidth]{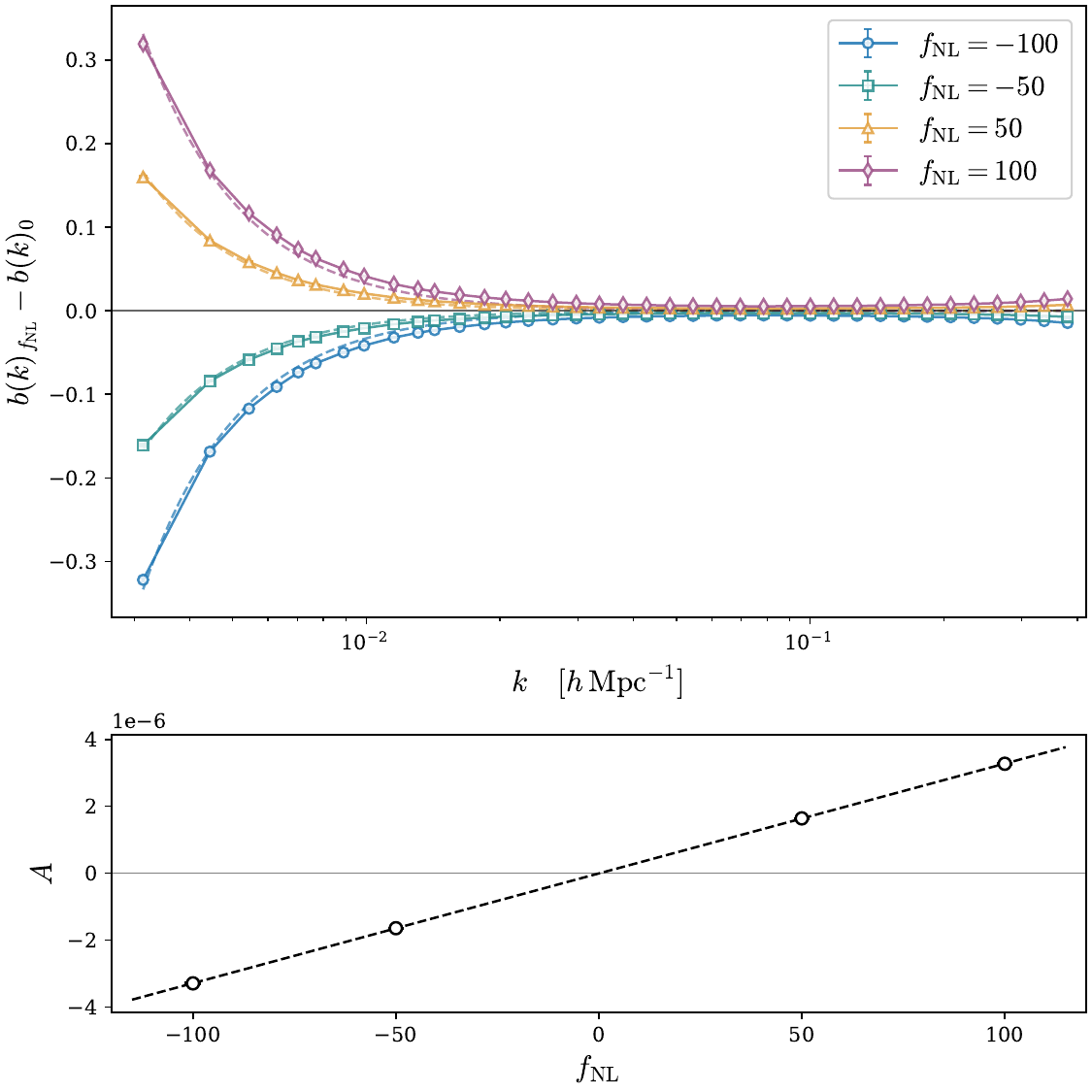}
  \end{tabular}
  \caption{Recovery of the large-scale bias signature of local primordial
non-Gaussianity using ALPT dark-matter fields.
The test adopts a power-law tracer-bias prescription,
$\rho_{\rm tr}\propto(1+\delta_{\rm m})^\alpha$, with $\alpha=2$,
at $z=1.1$, using ten realisations in a periodic box of side length
$L=2000\,h^{-1}\,{\rm Mpc}$ on a $256^3$ mesh.
\textit{Top:} Scale-dependent bias difference,
$\Delta b(k)=b(k;f_{\rm NL})-b(k;0)$, for
$f_{\rm NL}=-100,-50,50$, and $100$.
The measurements recover the characteristic Dalal et al.~\citep{Dalal_2008}
$\Delta b(k)\propto f_{\rm NL}k^{-2}$ behaviour towards small
wavenumbers, with positive and negative $f_{\rm NL}$ producing
an enhancement and suppression, respectively.
Dashed curves show the corresponding $A/k^2$ fits.
\textit{Bottom:} Fitted amplitude $A$ as a function of $f_{\rm NL}$,
showing an approximately linear response. This fixed-redshift test isolates the leading PNG bias signature; its extension to evolving lightcones is discussed in Sec.~\ref{sec:webon:png}.
}
\label{fig:dalal_probe}
\end{figure}

We have implemented local primordial non-Gaussianity directly
in the initial conditions of \texttt{WebON}, allowing ALPT
dark-matter fields to be generated for different values of
$f_{\rm NL}$. Figure~\ref{fig:dalal_probe} presents a controlled
test at $z=1.1$, using ten realisations on a $256^3$ mesh in a
box of side length $2000\,h^{-1}\,{\rm Mpc}$.
For a power-law tracer-bias prescription with exponent
$\alpha=2$, the bias difference relative to the Gaussian case
recovers the expected large-scale enhancement or suppression.
Its amplitude varies approximately linearly with $f_{\rm NL}$ over the range tested: the upper panel of Fig.~\ref{fig:dalal_probe} displays the scale dependence, and the lower panel summarises the fitted amplitude. The $A/k^2$ fit tests the asymptotic shape; agreement with the full prediction in Eq.~\eqref{eq:png_bias_general} additionally requires the transfer function and a calibrated tracer response $b_\Phi$. In particular, the illustrative $\alpha=2$ mapping does not by itself impose the universal halo relation for $b_\Phi$.

The same leading large-scale signature can also be represented
within our environmental bias framework through the nonlocal tensor introduced in Eq.~\eqref{eq:png_hessian}, here written explicitly in real and Fourier space,
\begin{equation}
    \mathcal{H}_{ij}(\boldsymbol{x})
    \equiv \partial_i\partial_j\nabla^{-4}
    \delta_{\rm m}(\boldsymbol{x}),
    \qquad
    \widehat{\mathcal{H}}_{ij}(\boldsymbol{k})
    =-\frac{k_i k_j}{k^4}\,
    \widehat{\delta}_{\rm m}(\boldsymbol{k}),
    \quad k>0.
    \label{eq:png_environment_tensor}
\end{equation}
Although the scalar field entering the Hessian carries a
$k^{-4}$ weighting, the two spatial derivatives yield a tensor
with an overall $k^{-2}$ dependence. In particular,
$\mathrm{Tr}\,\mathcal{H}=\nabla^{-2}\delta_{\rm m}$,
which provides a potential-like environmental variable.
We classify regions using the eigenvalues of
$\mathcal{H}_{ij}$ and assign different bias weights to the
resulting environments. Suitable choices of these weights
reproduce the leading large-scale response, providing an
effective environmental representation of the primordial
non-Gaussian bias signal.

The PNG initial-condition test in Fig.~\ref{fig:dalal_probe} and the complementary environmental construction described above address the same leading bias signature through different routes. The environmental construction can capture the characteristic scale dependence and its amplitude response, as tested in the companion analysis; the figure shown here specifically concerns the ALPT initial-condition test. Equivalence to non-Gaussian initial
conditions beyond this bias statistic, including higher-order
correlations and tracer abundances, requires separate validation.

\paragraph{Scope of the validation.}
\label{sec:validation:scope}

The tests above assess the construction and redshift evolution of
the lightcone matter and web fields, together with controlled checks
of the optional GR and PNG modules. They are complemented by the
independent Abacus matter-field benchmark of ridged LPT
\citep{2026arXiv260313106K} and the published calibration of
hierarchical halo bias on coarse ALPT fields \citep{Coloma_2024}.
They do not, by themselves, validate every tracer population or
observable in the final \texttt{HOLI} mocks: tracer-dependent
redshift-space, bispectrum, stochasticity, exclusion, and
property-selected statistics require their own tests (Forero-S\'anchez
et al. and Tang et al., in preparation).

\section{Conclusions}
\label{sec:conclusions}

We have presented \texttt{WebON}, a fast method for mapping the
evolving dark-matter cosmic web directly onto the observer's past
lightcone, and described its role as the gravity backbone of the
\texttt{HOLI} mock strategy.

\texttt{WebON} assigns particles an emission redshift using a
Zel'dovich estimate of their Eulerian distance and evaluates their
displacements through a continuous-redshift correction to a limited
set of stored approximate-gravity fields. This avoids both the
replication of smaller periodic boxes and the stitching of
independently evolved snapshot shells. The method supports regularised
ALPT and eALPT displacement fields, with an optional ridging
completion for enhanced short-range morphology.

We have illustrated the coherent evolution of matter and
cosmic-web fields across the lightcone. The lightcone maps (Fig.~\ref{fig:lightcone_maps}), spectral web fields (Figs.~\ref{fig:tweb_minus2} and \ref{fig:tweb_zero}), and angular-clustering measurements (Fig.~\ref{fig:angular_power_spectrum}) provide complementary tests of the internal consistency of the \texttt{WebON} construction. Since the angular-clustering reference uses the same approximate
gravity model, these tests establish the internal consistency of
lightcone evolution, not the accuracy of the raw ALPT matter spectrum.
Independent Abacus comparisons for ridged LPT
\citep{2026arXiv260313106K} and the published percent-level
accuracy of halo power spectra recovered through \texttt{HICOBIAN}
from coarse ALPT fields \citep{Coloma_2024} already address
different components of the accuracy question. Optional ultra-large-scale
modules include an approximate relativistic position correction and
initial conditions with local primordial non-Gaussianity. The GR test (Fig.~\ref{fig:gr_probe}) exhibits the expected infrared enhancement of the corrected density field, while the PNG test (Fig.~\ref{fig:dalal_probe}) recovers the characteristic Dalal et al. scale-dependent bias and its approximately linear response to $f_{\rm NL}$. Future studies will follow this PNG response continuously along the lightcone, including its interplay with evolving tracer bias, selection, and relativistic corrections.

The \texttt{HOLI} strategy uses the spectral hierarchy of the cosmic
web as the interface between the resolved matter lightcone and the
unresolved tracer-formation model. Crucially, the approximate ALPT
lightcone fields can undergo post-processing ridging and then be
populated using calibrated hierarchical field-level bias, stochastic
sampling, and subgrid phase-space prescriptions. This modular
calibration allows accurate tracer clustering even where the raw
matter solver lacks comparable small-scale power. Existing work
establishes better than $2\%$ halo power-spectrum accuracy up to
$k\simeq0.8\,h\,\mathrm{Mpc}^{-1}$ for one calibrated
\texttt{HICOBIAN} configuration \citep{Coloma_2024}; the exact
performance is tracer and calibration dependent. Companion studies
extend these prescriptions to further tracer populations and
statistics (Forero-S\'anchez et al. and Tang et al., in preparation).

This separation of responsibilities makes \texttt{WebON} a reusable
lightcone backbone: the same phase-coherent matter and web
realisations can support different tracer calibrations and analysis
requirements without regenerating the underlying lightcone.

Building on its application to DESI, we intend to extend the
\texttt{HOLI} mock programme to additional spectroscopic surveys,
including Euclid, the Subaru Prime Focus Spectrograph (PFS), and
the MUltiplexed Survey Telescope (MUST), among others
\citep{DESI2016Science,Euclid2025Overview,Takada2014PFS,Zhao2026MUST}.
The modular framework allows the matter-lightcone construction
to be combined with tracer-specific bias calibrations, subgrid
phase-space prescriptions, and survey-dependent selection functions.
Our aim is to provide large ensembles of mock catalogues for
covariance estimation, validation of cosmological inference,
and forward modelling of observational systematics across
different tracer populations and redshift ranges.
Where survey volumes overlap, populations generated from common
matter realisations would also enable consistent tests of
cross-correlations and joint analyses.
These extensions will require dedicated calibration and validation
for the target populations, observables, and selection procedures
of each survey.

\section*{Acknowledgements}
FSK acknowledges the Instituto de Astrof\'isica de Canarias (IAC) for continuous support to the \textit{Cosmology with LSS probes} activities
and the Spanish Ministry of Innovation and Science / Agencia Estatal de Investigaci\'on for support of the project
\textit{Big Data of the Cosmic Web} (PID2020-120612GB-I00/AEI/10.13039/501100011033), under which this work has been conceived and carried out and  the \textit{FIRE (Field-level bayesian Inference to Reconstruct the univErse)} project, which has been recently funded by a \textit{Proyecto de Generaci\'on de Conocimiento 2024}, PID2024-160504NB-I00.
FS acknowledges support from the \textit{Institute for Fundamental Physics of the Universe} postdoctoral fellowship scheme. FS is also grateful to IAC for hospitality during part of the realization of this project.
KN acknowledges support from JSPS KAKENHI grants 24H00002, 24H00241, JP25K01032, and the  International Leading Research (ILR) project, JP22K21349.  KN also acknowledges support from the Kavli IPMU, the World Premier Research Center Initiative (WPI), UTIAS, and the University of Tokyo.
The authors thank Arnaud De Mattia for providing the \texttt{JAX} version of the \texttt{WebON} code. 
The authors acknowledge Light Bridges, S.L., in collaboration with Bechtle S.L. and the b-HAIS Center of Excellence in Tenerife, Canary Islands (Spain), for providing access to Indefeasible Computer Rights (ICR-DEMO 2025–2026) supporting the development and testing of \texttt{WebON} and \texttt{BRIDGE} codes.

\section*{Disclaimer}
The \texttt{WebON} code used in this work has been mainly written by FSK in C++ and was coded without using AI.

\texttt{WebON} is a modular framework whose methodological components are presented across three companion papers: the present work describing the lightcone evolution mode within the \texttt{HOLI} pipeline {to produce the DESI covariance mocks used for the DR2 Key Projects} (see: \url{www.cosmic-signal.org}); a paper on ridged Lagrangian perturbation theory \citep{2026arXiv260313106K}; a dedicated paper on the spectral hierarchical cosmic-web  \citep{2026arXiv260315834K}.
\texttt{WebON} has already been used in three previous publications: one presenting the \emph{Hierarchical cosmic web and assembly bias} \citep{Coloma_2024} approach for effective field-level bias modelling exploiting the hierarchical cosmic web $k^{-2}$ and $k^0$ levels; one presenting \emph{Cosmomia: cosmic-web based redshift-space halo distribution} \citep{Forero_2024} for subgrid modelling exploiting the knots $k^0$  level as attractors; and \emph{Fast and accurate Gaia-unWISE quasar mock catalogs from LPT and Eulerian bias} \citep{2026JCAP...06..037S}, using the  two-hundred 10 $h^{-1}\,\mathrm{Gpc}$ side  cubical volume simulations  performed with the lightcone evolution \textsc{WebOn} mode on the Leonardo@CINECA pre-exascale supercomputing facilities in September 2023 through the INAF\_C9A09 HPC proposal (637k core hrs). 
It has served as a reference for its \texttt{JAX} differentiable version in \emph{Differentiable Fuzzy Cosmic-Web for Field Level Inference} \citep[][]{2026A&A...709A.110R}, in particular the ridged n-LPT lightcone version. 
The \texttt{JAX} version of the \texttt{WebON} code used within DESI was developed by Arnaud De Mattia based on its C++ version.

\paragraph{Data availability.}
Data products generated in this work (intermediate lightcone fields and mock catalogues) are available from the corresponding author upon reasonable request.

\paragraph{Code availability.}
The \texttt{WebON}  and \texttt{HOLI} pipeline components used in this analysis are available upon reasonable request, subject to DESI collaboration policies.

\section*{Affiliations}

\noindent
$^{1}$ Instituto de Astrof\'{\i}sica de Canarias, V\'{\i}a L\'actea s/n, E-38205 La Laguna, Tenerife, Spain

\noindent
$^{2}$ Departamento de Astrof\'{\i}sica, Universidad de La Laguna, E-38206 La Laguna, Tenerife, Spain

\noindent
$^{3}$ Institute for Fundamental Physics of the Universe, Via Beirut 2, I-34151 Trieste, Italy

\noindent
$^{4}$ SISSA -- International School for Advanced Studies, Via Bonomea 265, 34136 Trieste, Italy

\noindent
$^{5}$ INAF -- Osservatorio Astronomico di Trieste, Via G. B. Tiepolo 11, I-34131 Trieste, Italy

\noindent
$^{6}$ INFN -- National Institute for Nuclear Physics, Via Valerio 2, I-34127 Trieste, Italy

\noindent
$^{7}$ Institut de Ci\`encies del Cosmos (ICCUB), Universitat de Barcelona (UB), c. Mart\'i i Franqu\`es 1, E-08028 Barcelona, Spain

\noindent
$^{8}$ High Energy Physics Division, Argonne National Laboratory, Lemont, IL 60439, USA

\noindent
$^{9}$ Observatorio Astron\'omico Nacional, Universidad Nacional de Colombia, Av. Carrera 30 No.~45-03, 111321, Bogot\'a, Colombia

\noindent
$^{10}$ Department of Astronomy, Tsinghua University, Beijing 100084, China

\noindent
$^{11}$ Institute of Space Sciences (ICE, CSIC), Campus UAB, Carrer de Can Magrans, s/n, 08193 Barcelona, Spain

\noindent
$^{12}$ Theoretical Astrophysics, Department of Earth and Space Science, Graduate School of Science, The University of Osaka, 1-1 Machikaneyama, Toyonaka, Osaka 560-0043, Japan

\noindent
$^{13}$ Theoretical Joint Research, Forefront Research Center, The University of Osaka, Toyonaka, Osaka 560-0043, Japan

\noindent
$^{14}$ Kavli IPMU (WPI), UTIAS, The University of Tokyo, Kashiwa, Chiba 277-8583, Japan

\noindent
$^{15}$ Department of Physics and Astronomy, University of Nevada, Las Vegas, 4505 S. Maryland Pkwy, Las Vegas, NV 89154-4002, USA

\noindent
$^{16}$ Nevada Center for Astrophysics, University of Nevada, Las Vegas, 4505 S. Maryland Pkwy, Las Vegas, NV 89154-4002, USA

\bibliographystyle{JHEP}
\bibliography{references}

\providecommand{\href}[2]{#2}\begingroup\raggedright\begin{thebibliography}{10}

\bibitem{2009LNP...665..291V}
R.~{van de Weygaert} and W.~{Schaap}, \emph{{The Cosmic Web: Geometric Analysis}},  in \emph{Data Analysis in Cosmology}, V.J.~{Mart{\'\i}nez}, E.~{Saar}, E.~{Mart{\'\i}nez-Gonz{\'a}lez} and M.J.~{Pons-Border{\'\i}a}, eds., vol.~665, pp.~291--413 (2009), \href{https://doi.org/10.1007/978-3-540-44767-2_11}{DOI}.

\bibitem{2015ApJ...806..113G}
R.~{G{\'e}nova-Santos}, F.~{Atrio-Barandela}, F.S.~{Kitaura} and J.P.~{M{\"u}cket}, \emph{{Constraining the Baryon Fraction in the Warm Hot Intergalactic Medium at Low Redshifts with Planck Data}}, \href{https://doi.org/10.1088/0004-637X/806/1/113}{\emph{\apj} {\bfseries 806} (2015) 113} [\href{https://arxiv.org/abs/1501.01445}{{\ttfamily 1501.01445}}].

\bibitem{2015Natur.528..105E}
D.~{Eckert}, M.~{Jauzac}, H.~{Shan}, J.-P.~{Kneib}, T.~{Erben}, H.~{Israel} et~al., \emph{{Warm-hot baryons comprise 5-10 per cent of filaments in the cosmic web}}, \href{https://doi.org/10.1038/nature16058}{\emph{\nat} {\bfseries 528} (2015) 105} [\href{https://arxiv.org/abs/1512.00454}{{\ttfamily 1512.00454}}].

\bibitem{2019A&A...624A..48D}
A.~{de Graaff}, Y.-C.~{Cai}, C.~{Heymans} and J.A.~{Peacock}, \emph{{Probing the missing baryons with the Sunyaev-Zel'dovich effect from filaments}}, \href{https://doi.org/10.1051/0004-6361/201935159}{\emph{\aap} {\bfseries 624} (2019) A48} [\href{https://arxiv.org/abs/1709.10378}{{\ttfamily 1709.10378}}].

\bibitem{2016PhRvL.116q1301K}
F.-S.~{Kitaura}, C.-H.~{Chuang}, Y.~{Liang}, C.~{Zhao}, C.~{Tao}, S.~{Rodr{\'\i}guez-Torres} et~al., \emph{{Signatures of the Primordial Universe from Its Emptiness: Measurement of Baryon Acoustic Oscillations from Minima of the Density Field}}, \href{https://doi.org/10.1103/PhysRevLett.116.171301}{\emph{\prl} {\bfseries 116} (2016) 171301} [\href{https://arxiv.org/abs/1511.04405}{{\ttfamily 1511.04405}}].

\bibitem{2016MNRAS.459.4020L}
Y.~{Liang}, C.~{Zhao}, C.-H.~{Chuang}, F.-S.~{Kitaura} and C.~{Tao}, \emph{{Measuring baryon acoustic oscillations from the clustering of voids}}, \href{https://doi.org/10.1093/mnras/stw884}{\emph{\mnras} {\bfseries 459} (2016) 4020} [\href{https://arxiv.org/abs/1511.04391}{{\ttfamily 1511.04391}}].

\bibitem{2020MNRAS.491.4554Z}
C.~{Zhao}, C.-H.~{Chuang}, F.-S.~{Kitaura}, Y.~{Liang}, M.~{Pellejero-Ibanez}, C.~{Tao} et~al., \emph{{Improving baryon acoustic oscillation measurement with the combination of cosmic voids and galaxies}}, \href{https://doi.org/10.1093/mnras/stz3339}{\emph{\mnras} {\bfseries 491} (2020) 4554} [\href{https://arxiv.org/abs/1802.03990}{{\ttfamily 1802.03990}}].

\bibitem{2017PhRvD..95f3528C}
C.-H.~{Chuang}, F.-S.~{Kitaura}, Y.~{Liang}, A.~{Font-Ribera}, C.~{Zhao}, P.~{McDonald} et~al., \emph{{Linear redshift space distortions for cosmic voids based on galaxies in redshift space}}, \href{https://doi.org/10.1103/PhysRevD.95.063528}{\emph{\prd} {\bfseries 95} (2017) 063528} [\href{https://arxiv.org/abs/1605.05352}{{\ttfamily 1605.05352}}].

\bibitem{nadathur18}
S.~Nadathur, P.~Carter and W.J.~Percival, \emph{{A Zel'dovich reconstruction method for measuring redshift space distortions using cosmic voids}}, \href{https://doi.org/10.1093/mnras/sty2799}{\emph{Monthly Notices of the Royal Astronomical Society} {\bfseries 482} (2018) 2459} [\href{https://arxiv.org/abs/https://academic.oup.com/mnras/article-pdf/482/2/2459/26576217/sty2799.pdf}{{\ttfamily https://academic.oup.com/mnras/article-pdf/482/2/2459/26576217/sty2799.pdf}}].

\bibitem{2007PhRvL..98h1301P}
D.~{Park} and J.~{Lee}, \emph{{Void Ellipticity Distribution as a Probe of Cosmology}}, \href{https://doi.org/10.1103/PhysRevLett.98.081301}{\emph{\prl} {\bfseries 98} (2007) 081301} [\href{https://arxiv.org/abs/astro-ph/0610520}{{\ttfamily astro-ph/0610520}}].

\bibitem{2010MNRAS.403.1392L}
G.~{Lavaux} and B.D.~{Wandelt}, \emph{{Precision cosmology with voids: definition, methods, dynamics}}, \href{https://doi.org/10.1111/j.1365-2966.2010.16197.x}{\emph{\mnras} {\bfseries 403} (2010) 1392} [\href{https://arxiv.org/abs/0906.4101}{{\ttfamily 0906.4101}}].

\bibitem{2012MNRAS.426..440B}
E.G.P.~{Bos}, R.~{van de Weygaert}, K.~{Dolag} and V.~{Pettorino}, \emph{{The darkness that shaped the void: dark energy and cosmic voids}}, \href{https://doi.org/10.1111/j.1365-2966.2012.21478.x}{\emph{\mnras} {\bfseries 426} (2012) 440} [\href{https://arxiv.org/abs/1205.4238}{{\ttfamily 1205.4238}}].

\bibitem{2015PhRvD..92h3531P}
A.~{Pisani}, P.M.~{Sutter}, N.~{Hamaus}, E.~{Alizadeh}, R.~{Biswas}, B.D.~{Wandelt} et~al., \emph{{Counting voids to probe dark energy}}, \href{https://doi.org/10.1103/PhysRevD.92.083531}{\emph{\prd} {\bfseries 92} (2015) 083531} [\href{https://arxiv.org/abs/1503.07690}{{\ttfamily 1503.07690}}].

\bibitem{2019MNRAS.484.5267K}
A.~{Kov{\'a}cs}, C.~{S{\'a}nchez}, J.~{Garc{\'\i}a-Bellido}, J.~{Elvin-Poole}, N.~{Hamaus}, V.~{Miranda} et~al., \emph{{More out of less: an excess integrated Sachs-Wolfe signal from supervoids mapped out by the Dark Energy Survey}}, \href{https://doi.org/10.1093/mnras/stz341}{\emph{\mnras} {\bfseries 484} (2019) 5267} [\href{https://arxiv.org/abs/1811.07812}{{\ttfamily 1811.07812}}].

\bibitem{2001ApJ...557..117B}
M.L.~{Balogh}, D.~{Christlein}, A.I.~{Zabludoff} and D.~{Zaritsky}, \emph{{The Environmental Dependence of the Infrared Luminosity and Stellar Mass Functions}}, \href{https://doi.org/10.1086/321670}{\emph{\apj} {\bfseries 557} (2001) 117} [\href{https://arxiv.org/abs/astro-ph/0104042}{{\ttfamily astro-ph/0104042}}].

\bibitem{2004ApJ...615L.101B}
M.L.~{Balogh}, I.K.~{Baldry}, R.~{Nichol}, C.~{Miller}, R.~{Bower} and K.~{Glazebrook}, \emph{{The Bimodal Galaxy Color Distribution: Dependence on Luminosity and Environment}}, \href{https://doi.org/10.1086/426079}{\emph{\apjl} {\bfseries 615} (2004) L101} [\href{https://arxiv.org/abs/astro-ph/0406266}{{\ttfamily astro-ph/0406266}}].

\bibitem{2004MNRAS.353..713K}
G.~{Kauffmann}, S.D.M.~{White}, T.M.~{Heckman}, B.~{M{\'e}nard}, J.~{Brinchmann}, S.~{Charlot} et~al., \emph{{The environmental dependence of the relations between stellar mass, structure, star formation and nuclear activity in galaxies}}, \href{https://doi.org/10.1111/j.1365-2966.2004.08117.x}{\emph{\mnras} {\bfseries 353} (2004) 713} [\href{https://arxiv.org/abs/astro-ph/0402030}{{\ttfamily astro-ph/0402030}}].

\bibitem{2005ApJ...629..143B}
M.R.~{Blanton}, D.~{Eisenstein}, D.W.~{Hogg}, D.J.~{Schlegel} and J.~{Brinkmann}, \emph{{Relationship between Environment and the Broadband Optical Properties of Galaxies in the Sloan Digital Sky Survey}}, \href{https://doi.org/10.1086/422897}{\emph{\apj} {\bfseries 629} (2005) 143} [\href{https://arxiv.org/abs/astro-ph/0310453}{{\ttfamily astro-ph/0310453}}].

\bibitem{2006MNRAS.373..469B}
I.K.~{Baldry}, M.L.~{Balogh}, R.G.~{Bower}, K.~{Glazebrook}, R.C.~{Nichol}, S.P.~{Bamford} et~al., \emph{{Galaxy bimodality versus stellar mass and environment}}, \href{https://doi.org/10.1111/j.1365-2966.2006.11081.x}{\emph{\mnras} {\bfseries 373} (2006) 469} [\href{https://arxiv.org/abs/astro-ph/0607648}{{\ttfamily astro-ph/0607648}}].

\bibitem{2007ApJ...655L...5A}
M.A.~{Arag{\'o}n-Calvo}, R.~{van de Weygaert}, B.J.T.~{Jones} and J.M.~{van der Hulst}, \emph{{Spin Alignment of Dark Matter Halos in Filaments and Walls}}, \href{https://doi.org/10.1086/511633}{\emph{\apjl} {\bfseries 655} (2007) L5} [\href{https://arxiv.org/abs/astro-ph/0610249}{{\ttfamily astro-ph/0610249}}].

\bibitem{2008ApJ...688...78L}
J.~{Lee} and B.~{Lee}, \emph{{The Variation of Galaxy Morphological Type with Environmental Shear}}, \href{https://doi.org/10.1086/592181}{\emph{\apj} {\bfseries 688} (2008) 78} [\href{https://arxiv.org/abs/0801.1558}{{\ttfamily 0801.1558}}].

\bibitem{2008MNRAS.383..907B}
N.M.~{Ball}, J.~{Loveday} and R.J.~{Brunner}, \emph{{Galaxy colour, morphology and environment in the Sloan Digital Sky Survey}}, \href{https://doi.org/10.1111/j.1365-2966.2007.12627.x}{\emph{\mnras} {\bfseries 383} (2008) 907} [\href{https://arxiv.org/abs/astro-ph/0610171}{{\ttfamily astro-ph/0610171}}].

\bibitem{2012MNRAS.421L.137L}
N.I.~{Libeskind}, Y.~{Hoffman}, A.~{Knebe}, M.~{Steinmetz}, S.~{Gottl{\"o}ber}, O.~{Metuki} et~al., \emph{{The cosmic web and the orientation of angular momenta}}, \href{https://doi.org/10.1111/j.1745-3933.2012.01222.x}{\emph{\mnras} {\bfseries 421} (2012) L137} [\href{https://arxiv.org/abs/1201.3365}{{\ttfamily 1201.3365}}].

\bibitem{2013AJ....145..120B}
B.~{Beygu}, K.~{Kreckel}, R.~{van de Weygaert}, J.M.~{van der Hulst} and J.H.~{van Gorkom}, \emph{{An Interacting Galaxy System along a Filament in a Void}}, \href{https://doi.org/10.1088/0004-6256/145/5/120}{\emph{\aj} {\bfseries 145} (2013) 120} [\href{https://arxiv.org/abs/1303.0538}{{\ttfamily 1303.0538}}].

\bibitem{2014MNRAS.445..988N}
S.E.~{Nuza}, F.-S.~{Kitaura}, S.~{He{\ss}}, N.I.~{Libeskind} and V.~{M{\"u}ller}, \emph{{The cosmic web of the Local Universe: cosmic variance, matter content and its relation to galaxy morphology}}, \href{https://doi.org/10.1093/mnras/stu1746}{\emph{\mnras} {\bfseries 445} (2014) 988} [\href{https://arxiv.org/abs/1406.1004}{{\ttfamily 1406.1004}}].

\bibitem{2015ApJ...802...82F}
M.E.~{Filho}, J.~{S{\'a}nchez Almeida}, C.~{Mu{\~n}oz-Tu{\~n}{\'o}n}, S.E.~{Nuza}, F.~{Kitaura} and S.~{He{\ss}}, \emph{{Extremely Metal-poor Galaxies: The Environment}}, \href{https://doi.org/10.1088/0004-637X/802/2/82}{\emph{\apj} {\bfseries 802} (2015) 82} [\href{https://arxiv.org/abs/1501.06709}{{\ttfamily 1501.06709}}].

\bibitem{2017ApJ...848...60Y}
X.~{Yang}, Y.~{Zhang}, T.~{Lu}, H.~{Wang}, F.~{Shi}, D.~{Tweed} et~al., \emph{{Revealing the Cosmic Web-dependent Halo Bias}}, \href{https://doi.org/10.3847/1538-4357/aa8c7a}{\emph{\apj} {\bfseries 848} (2017) 60} [\href{https://arxiv.org/abs/1704.02451}{{\ttfamily 1704.02451}}].

\bibitem{2018MNRAS.473.2486S}
J.~{Shi} and R.K.~{Sheth}, \emph{{Dependence of halo bias on mass and environment}}, \href{https://doi.org/10.1093/mnras/stx2277}{\emph{\mnras} {\bfseries 473} (2018) 2486} [\href{https://arxiv.org/abs/1707.04096}{{\ttfamily 1707.04096}}].

\bibitem{2018MNRAS.473.3941F}
J.D.~{Fisher} and A.~{Faltenbacher}, \emph{{Cosmic web type dependence of halo clustering}}, \href{https://doi.org/10.1093/mnras/stx2496}{\emph{\mnras} {\bfseries 473} (2018) 3941}.

\bibitem{2019MNRAS.485.2367C}
W.~{Cui}, A.~{Knebe}, N.I.~{Libeskind}, S.~{Planelles}, X.~{Yang}, W.~{Cui} et~al., \emph{{The large-scale environment from cosmological simulations II: The redshift evolution and distributions of baryons}}, \href{https://doi.org/10.1093/mnras/stz565}{\emph{\mnras} {\bfseries 485} (2019) 2367} [\href{https://arxiv.org/abs/1902.09522}{{\ttfamily 1902.09522}}].

\bibitem{2019MNRAS.483.4501A}
S.~{Alam}, Y.~{Zu}, J.A.~{Peacock} and R.~{Mandelbaum}, \emph{{Cosmic web dependence of galaxy clustering and quenching in SDSS}}, \href{https://doi.org/10.1093/mnras/sty3477}{\emph{\mnras} {\bfseries 483} (2019) 4501} [\href{https://arxiv.org/abs/1801.04878}{{\ttfamily 1801.04878}}].

\bibitem{2019OJAp....2E...7A}
M.A.~{Aragon Calvo}, M.C.~{Neyrinck} and J.~{Silk}, \emph{{Galaxy Quenching from Cosmic Web Detachment}}, \href{https://doi.org/10.21105/astro.1607.07881}{\emph{The Open Journal of Astrophysics} {\bfseries 2} (2019) 7} [\href{https://arxiv.org/abs/1607.07881}{{\ttfamily 1607.07881}}].

\bibitem{2019MNRAS.482.1900H}
J.~{Han}, Y.~{Li}, Y.~{Jing}, T.~{Nishimichi}, W.~{Wang} and C.~{Jiang}, \emph{{The multidimensional dependence of halo bias in the eye of a machine: a tale of halo structure, assembly, and environment}}, \href{https://doi.org/10.1093/mnras/sty2822}{\emph{\mnras} {\bfseries 482} (2019) 1900}.

\bibitem{2019MNRAS.487.1607G}
P.~{Ganeshaiah Veena}, M.~{Cautun}, E.~{Tempel}, R.~{van de Weygaert} and C.S.~{Frenk}, \emph{{The Cosmic Ballet II: spin alignment of galaxies and haloes with large-scale filaments in the EAGLE simulation}}, \href{https://doi.org/10.1093/mnras/stz1343}{\emph{\mnras} {\bfseries 487} (2019) 1607} [\href{https://arxiv.org/abs/1903.06716}{{\ttfamily 1903.06716}}].

\bibitem{2018MNRAS.473.1195L}
N.I.~{Libeskind}, R.~{van de Weygaert}, M.~{Cautun}, B.~{Falck}, E.~{Tempel}, T.~{Abel} et~al., \emph{{Tracing the cosmic web}}, \href{https://doi.org/10.1093/mnras/stx1976}{\emph{\mnras} {\bfseries 473} (2018) 1195} [\href{https://arxiv.org/abs/1705.03021}{{\ttfamily 1705.03021}}].

\bibitem{2002PhR...367....1B}
F.~{Bernardeau}, S.~{Colombi}, E.~{Gazta{\~n}aga} and R.~{Scoccimarro}, \emph{{Large-scale structure of the Universe and cosmological perturbation theory}}, \href{https://doi.org/10.1016/S0370-1573(02)00135-7}{\emph{\physrep} {\bfseries 367} (2002) 1} [\href{https://arxiv.org/abs/astro-ph/0112551}{{\ttfamily astro-ph/0112551}}].

\bibitem{1970A&A.....5...84Z}
Y.B.~{Zeldovich}, \emph{{Reprint of 1970A\&amp;A.....5...84Z. Gravitational instability: an approximate theory for large density perturbations.}}, {\emph{\aap} {\bfseries 500} (1970) 13}.

\bibitem{Bond_1996nat}
J.R.~{Bond}, L.~{Kofman} and D.~{Pogosyan}, \emph{{How filaments of galaxies are woven into the cosmic web}}, \href{https://doi.org/10.1038/380603a0}{\emph{\nat} {\bfseries 380} (1996) 603} [\href{https://arxiv.org/abs/astro-ph/9512141}{{\ttfamily astro-ph/9512141}}].

\bibitem{2007MNRAS.375..489H}
O.~{Hahn}, C.~{Porciani}, C.M.~{Carollo} and A.~{Dekel}, \emph{{Properties of dark matter haloes in clusters, filaments, sheets and voids}}, \href{https://doi.org/10.1111/j.1365-2966.2006.11318.x}{\emph{\mnras} {\bfseries 375} (2007) 489} [\href{https://arxiv.org/abs/astro-ph/0610280}{{\ttfamily astro-ph/0610280}}].

\bibitem{2018PhR...733....1D}
V.~{Desjacques}, D.~{Jeong} and F.~{Schmidt}, \emph{{Large-scale galaxy bias}}, \href{https://doi.org/10.1016/j.physrep.2017.12.002}{\emph{\physrep} {\bfseries 733} (2018) 1} [\href{https://arxiv.org/abs/1611.09787}{{\ttfamily 1611.09787}}].

\bibitem{McDonald09}
P.~{McDonald} and A.~{Roy}, \emph{{Clustering of dark matter tracers: generalizing bias for the coming era of precision LSS}}, \href{https://doi.org/10.1088/1475-7516/2009/08/020}{\emph{\jcap} {\bfseries 2009} (2009) 020} [\href{https://arxiv.org/abs/0902.0991}{{\ttfamily 0902.0991}}].

\bibitem{2012PhRvD..85h3509C}
K.C.~{Chan}, R.~{Scoccimarro} and R.K.~{Sheth}, \emph{{Gravity and large-scale nonlocal bias}}, \href{https://doi.org/10.1103/PhysRevD.85.083509}{\emph{\prd} {\bfseries 85} (2012) 083509} [\href{https://arxiv.org/abs/1201.3614}{{\ttfamily 1201.3614}}].

\bibitem{Coloma_2024}
J.M.~{Coloma-Nadal}, F.S.~{Kitaura}, J.E.~{Garc{\'\i}a-Farieta}, F.~{Sinigaglia}, G.~{Favole} and D.~{Forero S{\'a}nchez}, \emph{{The Hierarchical Cosmic Web and Assembly Bias}}, \href{https://doi.org/10.48550/arXiv.2403.19337}{\emph{arXiv e-prints} (2024) arXiv:2403.19337} [\href{https://arxiv.org/abs/2403.19337}{{\ttfamily 2403.19337}}].

\bibitem{2026arXiv260315834K}
F.-S.~{Kitaura} and F.~{Sinigaglia}, \emph{{Spectral Hierarchy of the Cosmic Web}}, \href{https://doi.org/10.48550/arXiv.2603.15834}{\emph{arXiv e-prints} (2026) arXiv:2603.15834} [\href{https://arxiv.org/abs/2603.15834}{{\ttfamily 2603.15834}}].

\bibitem{García-Farieta2024}
J.E.~{Garc{\'\i}a-Farieta}, A.~{Balaguera-Antol{\'\i}nez} and F.-S.~{Kitaura}, \emph{{Fast simulation mapping: From standard to modified gravity cosmologies using the bias assignment method}}, \href{https://doi.org/10.1051/0004-6361/202450755}{\emph{\aap} {\bfseries 690} (2024) A27} [\href{https://arxiv.org/abs/2405.10319}{{\ttfamily 2405.10319}}].

\bibitem{2026arXiv260313106K}
F.-S.~{Kitaura} and F.~{Sinigaglia}, \emph{{Ridged Lagrangian Perturbation Theory (RLPT)}}, \href{https://doi.org/10.48550/arXiv.2603.13106}{\emph{arXiv e-prints} (2026) arXiv:2603.13106} [\href{https://arxiv.org/abs/2603.13106}{{\ttfamily 2603.13106}}].

\bibitem{Kitaura_2013}
F.-S.~Kitaura and S.~He{\ss}, \emph{Cosmological structure formation with augmented lagrangian perturbation theory}, \href{https://doi.org/10.1093/mnrasl/slt101}{\emph{Monthly Notices of the Royal Astronomical Society: Letters} {\bfseries 435} (2013) L78}.

\bibitem{Kitaura_2024}
F.S.~{Kitaura}, F.~{Sinigaglia}, A.~{Balaguera-Antol{\'\i}nez} and G.~{Favole}, \emph{{The cosmic web from perturbation theory}}, \href{https://doi.org/10.1051/0004-6361/202345876}{\emph{\aap} {\bfseries 683} (2024) A215} [\href{https://arxiv.org/abs/2301.03648}{{\ttfamily 2301.03648}}].

\bibitem{Abel_2012}
T.~{Abel}, O.~{Hahn} and R.~{Kaehler}, \emph{{Tracing the dark matter sheet in phase space}}, \href{https://doi.org/10.1111/j.1365-2966.2012.21754.x}{\emph{Monthly Notices of the Royal Astronomical Society} {\bfseries 427} (2012) 61} [\href{https://arxiv.org/abs/1111.3944}{{\ttfamily 1111.3944}}].

\bibitem{2016PhRvD..94h3511H}
O.~{Hahn} and A.~{Paranjape}, \emph{{General relativistic screening in cosmological simulations}}, \href{https://doi.org/10.1103/PhysRevD.94.083511}{\emph{\prd} {\bfseries 94} (2016) 083511} [\href{https://arxiv.org/abs/1602.07699}{{\ttfamily 1602.07699}}].

\bibitem{Dalal_2008}
N.~{Dalal}, M.~{White}, J.R.~{Bond} and A.~{Shirokov}, \emph{{Halo Assembly Bias in Hierarchical Structure Formation}}, \href{https://doi.org/10.1086/591512}{\emph{\apj} {\bfseries 687} (2008) 12} [\href{https://arxiv.org/abs/0803.3453}{{\ttfamily 0803.3453}}].

\bibitem{Peraza_2026}
K.~{Peraza Vera}, F.-S.~{Kitaura} and F.~{Sinigaglia}, \emph{{Primordial non-Gaussianity as a large-scale cosmic-web bias}}, \href{https://doi.org/10.48550/arXiv.2611.09362}{\emph{arXiv e-prints} (2026) arXiv:2611.09362} [\href{https://arxiv.org/abs/2611.09362}{{\ttfamily 2611.09362}}].

\bibitem{2026A&A...709A.110R}
P.~{Rossell{\'o}}, F.-S.~{Kitaura}, D.~{Forero S{\'a}nchez}, F.~{Sinigaglia} and G.~{Favole}, \emph{{Differentiable fuzzy cosmic web for field-level inference}}, \href{https://doi.org/10.1051/0004-6361/202555817}{\emph{\aap} {\bfseries 709} (2026) A110} [\href{https://arxiv.org/abs/2506.03969}{{\ttfamily 2506.03969}}].

\bibitem{Izard_2016}
A.~{Izard}, M.~{Crocce} and P.~{Fosalba}, \emph{{ICE-COLA: towards fast and accurate synthetic galaxy catalogues optimizing a quasi-N-body method}}, \href{https://doi.org/10.1093/mnras/stw797}{\emph{\mnras} {\bfseries 459} (2016) 2327} [\href{https://arxiv.org/abs/1509.04685}{{\ttfamily 1509.04685}}].

\bibitem{Ereza2024}
J.~{Ereza}, F.~{Prada}, A.~{Klypin}, T.~{Ishiyama}, A.~{Smith}, C.M.~{Baugh} et~al., \emph{{The Uchuu-GLAM BOSS and eBOSS LRG lightcones: Exploring clustering and covariance errors}}, \href{https://doi.org/10.48550/arXiv.2311.14456}{\emph{arXiv e-prints} (2023) arXiv:2311.14456} [\href{https://arxiv.org/abs/2311.14456}{{\ttfamily 2311.14456}}].

\bibitem{Variu_2024}
A.~{Variu}, S.~{Alam}, C.~{Zhao}, C.-H.~{Chuang}, Y.~{Yu}, D.~{Forero-S{\'a}nchez} et~al., \emph{{DESI mock challenge: constructing DESI galaxy catalogues based on FASTPM simulations}}, \href{https://doi.org/10.1093/mnras/stad3742}{\emph{\mnras} {\bfseries 527} (2024) 11539} [\href{https://arxiv.org/abs/2307.14197}{{\ttfamily 2307.14197}}].

\bibitem{2021MNRAS.504.5205C}
S.~{Contreras}, R.E.~{Angulo} and M.~{Zennaro}, \emph{{A flexible modelling of galaxy assembly bias}}, \href{https://doi.org/10.1093/mnras/stab1170}{\emph{\mnras} {\bfseries 504} (2021) 5205} [\href{https://arxiv.org/abs/2005.03672}{{\ttfamily 2005.03672}}].

\bibitem{2023MNRAS.525.3149C}
S.~{Contreras}, J.~{Chaves-Montero} and R.E.~{Angulo}, \emph{{Consistent clustering and lensing of SDSS-III BOSS galaxies with an extended abundance matching formalism}}, \href{https://doi.org/10.1093/mnras/stad2434}{\emph{\mnras} {\bfseries 525} (2023) 3149} [\href{https://arxiv.org/abs/2305.09637}{{\ttfamily 2305.09637}}].

\bibitem{2024A&A...689A..66O}
S.~{Ortega-Martinez}, S.~{Contreras} and R.~{Angulo}, \emph{{SHAMe-SF: Predicting the clustering of star-forming galaxies with an enhanced abundance matching model}}, \href{https://doi.org/10.1051/0004-6361/202449597}{\emph{\aap} {\bfseries 689} (2024) A66} [\href{https://arxiv.org/abs/2401.17374}{{\ttfamily 2401.17374}}].

\bibitem{2026ApJ..1002...90F}
G.~{Favole}, F.-S.~{Kitaura}, B.~{Hadzhiyska}, D.~{Eisenstein}, L.H.~{Garrison} and S.~{Bose}, \emph{{ELG{\texttimes}LRG Distribution through Dark Matter Halo Dynamics}}, \href{https://doi.org/10.3847/1538-4357/ae592a}{\emph{\apj} {\bfseries 1002} (2026) 90} [\href{https://arxiv.org/abs/2512.04362}{{\ttfamily 2512.04362}}].

\bibitem{Bond_1996b}
J.R.~{Bond} and S.T.~{Myers}, \emph{{The Peak-Patch Picture of Cosmic Catalogs. II. Validation}}, \href{https://doi.org/10.1086/192268}{\emph{\apjs} {\bfseries 103} (1996) 41}.

\bibitem{Bond_1996c}
J.R.~{Bond} and S.T.~{Myers}, \emph{{The Peak-Patch Picture of Cosmic Catalogs. III. Application to Clusters}}, \href{https://doi.org/10.1086/192269}{\emph{\apjs} {\bfseries 103} (1996) 63}.

\bibitem{Monaco_2002}
P.~Monaco, T.~Theuns, G.~Taffoni, F.~Governato, T.~Quinn and J.~Stadel, \emph{Predicting the number, spatial distribution, and merging history of dark matter halos}, \href{https://doi.org/10.1086/324182}{\emph{The Astrophysical Journal} {\bfseries 564} (2002) 8}.

\bibitem{Monaco_2013}
P.~{Monaco}, E.~{Sefusatti}, S.~{Borgani}, M.~{Crocce}, P.~{Fosalba}, R.K.~{Sheth} et~al., \emph{{An accurate tool for the fast generation of dark matter halo catalogues}}, \href{https://doi.org/10.1093/mnras/stt907}{\emph{\mnras} {\bfseries 433} (2013) 2389} [\href{https://arxiv.org/abs/1305.1505}{{\ttfamily 1305.1505}}].

\bibitem{Manera_2013}
M.~{Manera}, R.~{Scoccimarro}, W.J.~{Percival}, L.~{Samushia}, C.K.~{McBride}, A.J.~{Ross} et~al., \emph{{The clustering of galaxies in the SDSS-III Baryon Oscillation Spectroscopic Survey: a large sample of mock galaxy catalogues}}, \href{https://doi.org/10.1093/mnras/sts084}{\emph{\mnras} {\bfseries 428} (2013) 1036} [\href{https://arxiv.org/abs/1203.6609}{{\ttfamily 1203.6609}}].

\bibitem{Manera_2015}
M.~{Manera}, L.~{Samushia}, R.~{Tojeiro}, C.~{Howlett}, A.J.~{Ross}, W.J.~{Percival} et~al., \emph{{The clustering of galaxies in the SDSS-III Baryon Oscillation Spectroscopic Survey: mock galaxy catalogues for the low-redshift sample}}, \href{https://doi.org/10.1093/mnras/stu2465}{\emph{\mnras} {\bfseries 447} (2015) 437} [\href{https://arxiv.org/abs/1401.4171}{{\ttfamily 1401.4171}}].

\bibitem{Avila:2014nia}
S.~Avila, S.G.~Murray, A.~Knebe, C.~Power, A.S.G.~Robotham and J.~Garcia-Bellido, \emph{{HALOGEN: A tool for fast generation of mock halo catalogues}}, \href{https://doi.org/10.1093/mnras/stv711}{\emph{Mon. Not. Roy. Astron. Soc.} {\bfseries 450} (2015) 1856} [\href{https://arxiv.org/abs/1412.5228}{{\ttfamily 1412.5228}}].

\bibitem{Chuang_2015}
C.-H.~{Chuang}, C.~{Zhao}, F.~{Prada}, E.~{Munari}, S.~{Avila}, A.~{Izard} et~al., \emph{{nIFTy cosmology: Galaxy/halo mock catalogue comparison project on clustering statistics}}, \href{https://doi.org/10.1093/mnras/stv1289}{\emph{\mnras} {\bfseries 452} (2015) 686} [\href{https://arxiv.org/abs/1412.7729}{{\ttfamily 1412.7729}}].

\bibitem{Blot_2019}
L.~{Blot}, M.~{Crocce}, E.~{Sefusatti}, M.~{Lippich}, A.G.~{S{\'a}nchez}, M.~{Colavincenzo} et~al., \emph{{Comparing approximate methods for mock catalogues and covariance matrices II: power spectrum multipoles}}, \href{https://doi.org/10.1093/mnras/stz507}{\emph{\mnras} {\bfseries 485} (2019) 2806} [\href{https://arxiv.org/abs/1806.09497}{{\ttfamily 1806.09497}}].

\bibitem{Colavincenzo_2019}
M.~{Colavincenzo}, E.~{Sefusatti}, P.~{Monaco}, L.~{Blot}, M.~{Crocce}, M.~{Lippich} et~al., \emph{{Comparing approximate methods for mock catalogues and covariance matrices - III: bispectrum}}, \href{https://doi.org/10.1093/mnras/sty2964}{\emph{\mnras} {\bfseries 482} (2019) 4883} [\href{https://arxiv.org/abs/1806.09499}{{\ttfamily 1806.09499}}].

\bibitem{Lippich_2019}
M.~{Lippich}, A.G.~{S{\'a}nchez}, M.~{Colavincenzo}, E.~{Sefusatti}, P.~{Monaco}, L.~{Blot} et~al., \emph{{Comparing approximate methods for mock catalogues and covariance matrices - I. Correlation function}}, \href{https://doi.org/10.1093/mnras/sty2757}{\emph{\mnras} {\bfseries 482} (2019) 1786} [\href{https://arxiv.org/abs/1806.09477}{{\ttfamily 1806.09477}}].

\bibitem{Scoccimarro_2002}
R.~{Scoccimarro} and R.K.~{Sheth}, \emph{{PTHALOS: a fast method for generating mock galaxy distributions}}, \href{https://doi.org/10.1046/j.1365-8711.2002.04999.x}{\emph{\mnras} {\bfseries 329} (2002) 629} [\href{https://arxiv.org/abs/astro-ph/0106120}{{\ttfamily astro-ph/0106120}}].

\bibitem{Kitaura_2014}
F.-S.~Kitaura, G.~Yepes and F.~Prada, \emph{{Modelling baryon acoustic oscillations with perturbation theory and stochastic halo biasing}}, \href{https://doi.org/10.1093/mnrasl/slt172}{\emph{Monthly Notices of the Royal Astronomical Society: Letters} {\bfseries 439} (2013) L21} [\href{https://arxiv.org/abs/https://academic.oup.com/mnrasl/article-pdf/439/1/L21/3421996/slt172.pdf}{{\ttfamily https://academic.oup.com/mnrasl/article-pdf/439/1/L21/3421996/slt172.pdf}}].

\bibitem{Kitaura_2015}
F.-S.~{Kitaura}, H.~{Gil-Mar{\'\i}n}, C.G.~{Sc{\'o}ccola}, C.-H.~{Chuang}, V.~{M{\"u}ller}, G.~{Yepes} et~al., \emph{{Constraining the halo bispectrum in real and redshift space from perturbation theory and non-linear stochastic bias}}, \href{https://doi.org/10.1093/mnras/stv645}{\emph{\mnras} {\bfseries 450} (2015) 1836} [\href{https://arxiv.org/abs/1407.1236}{{\ttfamily 1407.1236}}].

\bibitem{Kitaura_2016}
F.-S.~{Kitaura}, S.~{Rodr{\'\i}guez-Torres}, C.-H.~{Chuang}, C.~{Zhao}, F.~{Prada}, H.~{Gil-Mar{\'\i}n} et~al., \emph{{The clustering of galaxies in the SDSS-III Baryon Oscillation Spectroscopic Survey: mock galaxy catalogues for the BOSS Final Data Release}}, \href{https://doi.org/10.1093/mnras/stv2826}{\emph{\mnras} {\bfseries 456} (2016) 4156} [\href{https://arxiv.org/abs/1509.06400}{{\ttfamily 1509.06400}}].

\bibitem{Zhao_2021}
C.~{Zhao}, C.-H.~{Chuang}, J.~{Bautista}, A.~{de Mattia}, A.~{Raichoor}, A.J.~{Ross} et~al., \emph{{The completed SDSS-IV extended Baryon Oscillation Spectroscopic Survey: 1000 multi-tracer mock catalogues with redshift evolution and systematics for galaxies and quasars of the final data release}}, \href{https://doi.org/10.1093/mnras/stab510}{\emph{\mnras} {\bfseries 503} (2021) 1149} [\href{https://arxiv.org/abs/2007.08997}{{\ttfamily 2007.08997}}].

\bibitem{Balaguera_2018}
A.~Balaguera-Antolínez, F.-S.~Kitaura, M.~Pellejero-Ibáñez, C.~Zhao and T.~Abel, \emph{{BAM: bias assignment method to generate mock catalogues}}, \href{https://doi.org/10.1093/mnrasl/sly220}{\emph{Monthly Notices of the Royal Astronomical Society: Letters} {\bfseries 483} (2018) L58} [\href{https://arxiv.org/abs/https://academic.oup.com/mnrasl/article-pdf/483/1/L58/54700852/mnrasl\_483\_1\_l58.pdf}{{\ttfamily https://academic.oup.com/mnrasl/article-pdf/483/1/L58/54700852/mnrasl\_483\_1\_l58.pdf}}].

\bibitem{Balaguera_2020}
A.~{Balaguera-Antol{\'\i}nez}, F.-S.~{Kitaura}, M.~{Pellejero-Ib{\'a}{\~n}ez}, M.~{Lippich}, C.~{Zhao}, A.G.~{S{\'a}nchez} et~al., \emph{{One simulation to have them all: performance of the Bias Assignment Method against N-body simulations}}, \href{https://doi.org/10.1093/mnras/stz3206}{\emph{\mnras} {\bfseries 491} (2020) 2565} [\href{https://arxiv.org/abs/1906.06109}{{\ttfamily 1906.06109}}].

\bibitem{Balaguera_2023}
A.~{Balaguera-Antol{\'\i}nez}, F.-S.~{Kitaura}, S.~{Alam}, C.-H.~{Chuang}, Y.~{Yu}, G.~{Favole} et~al., \emph{{DESI mock challenge. Halo and galaxy catalogues with the bias assignment method}}, \href{https://doi.org/10.1051/0004-6361/202245618}{\emph{\aap} {\bfseries 673} (2023) A130} [\href{https://arxiv.org/abs/2211.10640}{{\ttfamily 2211.10640}}].

\bibitem{Kitaura_2022}
F.S.~{Kitaura}, A.~{Balaguera-Antol{\'\i}nez}, F.~{Sinigaglia} and M.~{Pellejero-Ib{\'a}{\~n}ez}, \emph{{The cosmic web connection to the dark matter halo distribution through gravity}}, \href{https://doi.org/10.1093/mnras/stac671}{\emph{\mnras} {\bfseries 512} (2022) 2245} [\href{https://arxiv.org/abs/2005.11598}{{\ttfamily 2005.11598}}].

\bibitem{Sinigaglia_2024}
F.~{Sinigaglia}, F.S.~{Kitaura}, K.~{Nagamine}, Y.~{Oku} and A.~{Balaguera-Antol{\'\i}nez}, \emph{{Field-level Lyman-{\ensuremath{\alpha}} forest modeling in redshift space via augmented nonlocal Fluctuating Gunn-Peterson Approximation}}, \href{https://doi.org/10.1051/0004-6361/202346931}{\emph{\aap} {\bfseries 682} (2024) A21} [\href{https://arxiv.org/abs/2305.10428}{{\ttfamily 2305.10428}}].

\bibitem{2026JCAP...06..037S}
F.~{Sinigaglia}, F.-S.~{Kitaura}, M.~{Shiferaw}, G.~{Favole}, K.~{Storey-Fisher} and N.~{Arsenov}, \emph{{Fast and accurate Gaia-unWISE quasar mock catalogs from LPT and Eulerian bias}}, \href{https://doi.org/10.1088/1475-7516/2026/06/037}{\emph{\jcap} {\bfseries 2026} (2026) 037} [\href{https://arxiv.org/abs/2509.15890}{{\ttfamily 2509.15890}}].

\bibitem{Kitaura_2021}
F.-S.~{Kitaura}, M.~{Ata}, S.A.~{Rodr{\'\i}guez-Torres}, M.~{Hern{\'a}ndez-S{\'a}nchez}, A.~{Balaguera-Antol{\'\i}nez} and G.~{Yepes}, \emph{{COSMIC BIRTH: efficient Bayesian inference of the evolving cosmic web from galaxy surveys}}, \href{https://doi.org/10.1093/mnras/staa3774}{\emph{\mnras} {\bfseries 502} (2021) 3456} [\href{https://arxiv.org/abs/1911.00284}{{\ttfamily 1911.00284}}].

\bibitem{Vakili_2017}
M.~{Vakili}, F.-S.~{Kitaura}, Y.~{Feng}, G.~{Yepes}, C.~{Zhao}, C.-H.~{Chuang} et~al., \emph{{Accurate halo-galaxy mocks from automatic bias estimation and particle mesh gravity solvers}}, \href{https://doi.org/10.1093/mnras/stx2184}{\emph{\mnras} {\bfseries 472} (2017) 4144} [\href{https://arxiv.org/abs/1701.03765}{{\ttfamily 1701.03765}}].

\bibitem{Forero_2024}
D.~{Forero-S{\'a}nchez}, F.-S.~{Kitaura}, F.~{Sinigaglia}, J.~{Mar{\'\i}a Coloma-Nodal} and J.-P.~{Kneib}, \emph{{CosmoMIA: Cosmic Web-based redshift space halo distribution}}, \href{https://doi.org/10.48550/arXiv.2402.17581}{\emph{arXiv e-prints} (2024) arXiv:2402.17581} [\href{https://arxiv.org/abs/2402.17581}{{\ttfamily 2402.17581}}].

\bibitem{Zhao_2015}
C.~Zhao, F.-S.~Kitaura, C.-H.~Chuang, F.~Prada, G.~Yepes and C.~Tao, \emph{{Halo mass distribution reconstruction across the cosmic web}}, \href{https://doi.org/10.1093/mnras/stv1262}{\emph{Monthly Notices of the Royal Astronomical Society} {\bfseries 451} (2015) 4266} [\href{https://arxiv.org/abs/https://academic.oup.com/mnras/article-pdf/451/4/4266/3901283/stv1262.pdf}{{\ttfamily https://academic.oup.com/mnras/article-pdf/451/4/4266/3901283/stv1262.pdf}}].

\bibitem{Ramakrishnan_2021}
S.~{Ramakrishnan}, A.~{Paranjape} and R.K.~{Sheth}, \emph{{Mock halo catalogues: assigning unresolved halo properties using correlations with local halo environment}}, \href{https://doi.org/10.1093/mnras/stab541}{\emph{\mnras} {\bfseries 503} (2021) 2053} [\href{https://arxiv.org/abs/2012.10170}{{\ttfamily 2012.10170}}].

\bibitem{DESI2016Science}
{DESI Collaboration}, A.~Aghamousa et~al., \emph{The {DESI} experiment part {I}: Science, targeting, and survey design}, \href{https://doi.org/10.48550/arXiv.1611.00036}{\emph{arXiv e-prints} (2016) } [\href{https://arxiv.org/abs/1611.00036}{{\ttfamily 1611.00036}}].

\bibitem{Euclid2025Overview}
{Euclid Collaboration}, Y.~Mellier et~al., \emph{{Euclid}. {I}. overview of the {Euclid} mission}, \href{https://doi.org/10.1051/0004-6361/202450810}{\emph{Astronomy \& Astrophysics} {\bfseries 697} (2025) A1} [\href{https://arxiv.org/abs/2405.13491}{{\ttfamily 2405.13491}}].

\bibitem{Takada2014PFS}
M.~Takada, R.S.~Ellis, M.~Chiba et~al., \emph{Extragalactic science, cosmology, and galactic archaeology with the {Subaru Prime Focus Spectrograph}}, \href{https://doi.org/10.1093/pasj/pst019}{\emph{Publications of the Astronomical Society of Japan} {\bfseries 66} (2014) R1} [\href{https://arxiv.org/abs/1206.0737}{{\ttfamily 1206.0737}}].

\bibitem{Zhao2026MUST}
C.~Zhao, S.~Huang, M.~He et~al., \emph{{MUltiplexed Survey Telescope (MUST)} science white paper {I}: Overview of large-scale structure cosmology in the era of {Stage-V} spectroscopic surveys}, \href{https://doi.org/10.1007/s11433-026-2998-9}{\emph{Science China Physics, Mechanics \& Astronomy} {\bfseries 69} (2026) 119511} [\href{https://arxiv.org/abs/2411.07970}{{\ttfamily 2411.07970}}].

\end{thebibliography}\endgroup

\end{document}